\documentclass[11pt]{article}

\usepackage[truedimen,margin=25truemm]{geometry}

\usepackage{amsmath}

\DeclareMathAlphabet\mathbfcal{OMS}{cmsy}{b}{d}

\usepackage{special_def}

\title{Network Volume Anomaly Detection and Identification in\\ Large-scale Networks based on\\ Online Time-structured Traffic Tensor Tracking}

\author{Hiroyuki Kasai\thanks{H. Kasai is with the Graduate School of Informatics and Engineering, The University of Electro-Communications, 1-5-1 Chofugaoka, Chofu-shi, Tokyo, 182-8585, Japan (e-mail: kasai@is.uec.ac.jp, web:www.kasailab.com)} \and Wolfgang Kellerer\thanks{W. Kellerer is with the Department of Electrical and Computer Engineering, Technical University of Munich, Munich 80290, Germany (e-mail: wolfgang@tum.de, www.lkn.ei.tum.de)} \and Martin Kleinsteuber\thanks{M. Kleinsteuber is with the Department of Electrical and Computer Engineering, Technical University of Munich, Munich 80290, Germany (e-mail: kleinsteuber@tum.de, www.gol.ei.tum.de)}}

\begin{document}

\maketitle

\begin{abstract}
This paper addresses network anomography, that is, the problem of inferring network-level anomalies from indirect link measurements. This problem is cast as a low-rank subspace tracking problem for normal flows under incomplete observations, and an outlier detection problem for abnormal flows. Since traffic data is large-scale time-structured data accompanied with noise and outliers under partial observations, an efficient modeling method is essential. To this end, this paper proposes an online subspace tracking of a Hankelized time-structured traffic tensor for normal flows based on the Candecomp/PARAFAC decomposition exploiting the recursive least squares (RLS) algorithm. We estimate abnormal flows as outlier sparse flows via sparsity maximization in the underlying under-constrained linear-inverse problem. A major advantage is that our algorithm estimates normal flows by low-dimensional matrices with time-directional features as well as the spatial correlation of multiple links without using the past observed measurements and the past model parameters. Extensive numerical evaluations show that the proposed algorithm achieves faster convergence per iteration of model approximation, and better volume anomaly detection performance compared to state-of-the-art algorithms. 
\end{abstract}

\section{Introduction}
Diagnosing unusual events (called ``anomalies") in a large-scale network like Internet Service Providers and enterprise networks is critical and challenging for both network operators and end users. Anomalies occur due to activity from malicious operations, or misconfigurations and failures of network equipments. This paper addresses ``{\it traffic volume anomaly}", which means large and sudden positive or negative {\it traffic volume changes} due to strong variances in traffic flows. These changes are typically caused by unexpected events such as alpha events (e.g., large file transfers), outages coming from network equipment failures, network attacks like denial-of-service attacks (DoS), and traffic shifts. Diagnosing such a volume anomaly in a flow consists of three steps; detection, identification and quantification \cite{Lakhina_SIGCOMM_2004}. The {\it detection} step is to unveil time points when the network is facing an anomaly. The {\it identification} step consists of selecting the right anomaly type from a set of possible candidate anomalies. The identification step addressed in this paper additionally allows to identify which flow experiences such an anomaly. Finally, the {\it quantification} step is about to estimate the number of additional or missing bytes in the underlying traffic flow. The detection and identification problems are of interest in this paper. The quantification problem is essential but extremely challenging in the online-based approach under an incomplete observation situation that this paper addresses, and this remains an open problem.

Network-wide traffic is typically expressed in the form of matrices or multidimensional arrays, i.e., tensors. In general, a traffic volume exchanged between every pair of an ingress and an egress node or PoP (Point of Presence) during a given time period forms a two-dimensional non-negative matrix, often called the {\it traffic matrix} or the {\it flow matrix}. This paper explicitly uses the term flow matrix to  avoid confusion with the link matrix mentioned hereafter. A flow is referred to as an {\it origin-destination} (OD) flow. Here, the term {\it traffic volume} refers to the  number of bytes, packets, or flows measured in a certain time interval at one point in the network. A noteworthy point is that one single OD flow traverses multiple links based on the routing tables, thereby a volume anomaly in one single OD flow is visible across several links simultaneously. For diagnosing, a large-scale direct collecting of {\it flow measurements}, which is {\it flow-level data}, is extremely resource intensive due to the collection of {\it fine-grained} data. It also requires a flow monitoring infrastructure across an entire network, and this is extremely costly. To the contrary, {\it link measurements}, which are {\it device-level data}, can be easily collected by the Simple Network Management Protocol (SNMP) that periodically collects device readings, but provides only {\it coarse-grained} information. The link measurements are generally expressed as the {\it link matrix}, which represents traffic volume of each link over time, and the link matrix is obtained by multiplying the {\it routing matrix} to the flow matrix. Practical limitations are that the collected link measurements are often noisy \cite{Roughan_JECE_2010}, contain {\it missing data} due to using the unreliable UDP transport, and lack measurement synchronization across an entire network. It should also be noted that the link measurements are a linear combination of OD flows, and, the observed traffic on each link is the {\it superposition} of multiple OD flows. 

Consequently, this paper addresses the {\it flow matrix estimation problem} from the link matrix, which is referred to as {\it network tomography} \cite{Vardi_JASA_1996,Zhang_ACM_IMC_2005}, to diagnose traffic volume anomaly. This estimates a {\it directly unobservable} flow matrix from the directly observable link matrix by using standard SNMP per-link byte counts. We also refer to the problem of inferring anomalies from these indirect measurement of the flow matrix as {\it network anomography}. It should be noted that this is different from and more complex than network tomography, in the way that network anomography is performed against sequential measurements over a certain period of time, rather than from a single snapshot of measurements.
Despite extensive studies for decades in this field, network tomography and anomography represent key technical issues of network management facing network evolution. 

The classical but key challenges that lie at the core of flow matrix estimation and volume anomaly diagnosis are as follows; the volume anomaly diagnosis problem stems from the fact that it only uses the link measurements, whereas the number of links in a network is generally much smaller than that of OD flows \cite{Lakhina_SIGCOMM_2004}. The problem is described as an {\it under-constrained linear-inverse problem}, where the solution relies on a prior model of the flow matrix (e.g., the Poisson model \cite{Vardi_JASA_1996}, the gravity model \cite{Zhang_SIGMETRICS_2003,Zhang_IEEEACMTranNW_2005}). 
An additional challenge comes from the superimposed OD flows on a link. While an OD flow has pronounced spikes, the spikes are dwarfed in the corresponding link traffic. In other words, clearly recognizable anomalous spikes in the flow matrix can often be covered in the link matrix through ``dreadful interference" of the superimposed OD flows. 

Keeping these challenges in our mind, we further need to consider an efficient and robust way to handle noisy, incomplete, high-dimensional flow matrices when the network size becomes larger. 
To this end, our motivation and approach for the primary contribution of this paper are summarized as follows;
considering that the flow and link matrices have time-directional structure such as periodicity and seasonality, a novel analysis method uncovering latent time-directional structure inside matrices is required. For this purpose, a {\it Hankel matrix} representation of the link matrix with the {\it tensor structure} plays a crucial role. This paper specifically transfers this link matrix to the {\it Hankelized time-structured traffic tensor}, and analyzes it. Moreover, since these matrices have high-dimensional data  accompanied with noise, a {\it subspace-based approach}, i.e., a low-rank approximation approach, is desirable to robustly model underlying latent behavior of network traffic. Thereby, we exploit a low-rank {\it tensor decomposition} not only to model robustly normal flows but also to compute big link matrices more efficiently by {\it thin} matrices rather than an entire size of matrices. Besides, as for the detection of abnormal flows, this can be newly cast as an outlier detection problem via the {\it sparsity constraint} formulation. Furthermore, the flow matrix estimation requires tolerance and insensitivity to missing measurements and non-synchronized measurements due to the unreliable UDP transport, network-wide monitoring, or machine failures. This requires to interpolate missing values in the link matrix \cite{Roughan_IEEEACM_TranNW_2012}, and motivates us to support a {\it matrix completion} function. Additionally, because the flow matrix and the link matrix grow infinitely as time goes by, an {\it online-based} approach without storing all the past measurements and model parameters is vital and effective. Finally, considering that the underlying subspace changes dramatically, and the processing speed is faster than the data acquiring speed,
this motivates us to adopt a {\it second-order optimization} algorithm because the fast convergence property in each iteration (analysis) is preferred over convergence in computation time. In fact, sampling of the link measurements is periodical but intermittent (e.g., every 5 minutes in basic SNMP).

Many efforts have been done so far in order to identify anomalies by analyzing network traffic \cite{Barford_ACM_IM_2002,Krishnamurthy_IMC_2003,Lakhina_SIGCOMM_2004,Soule_IMC_2005,Zhang_ACM_IMC_2005,Ringberg_ACMSigmetrics_2007,Roughan_IEEEACM_TranNW_2012}. They have attempt to expose anomalies by detecting deviations or errors from the constructed underlying model of normal traffic. There is, however, no single approach to satisfy the required algorithm capabilities that can handle noisy, high-dimensional, and time-series data with missing measurements in an online fashion. 

Consequently, this paper presents a new proposal for an online subspace tracking of the Hankelized time-structured traffic tensor for normal flows based on the Candecomp/PARAFAC tensor decomposition by exploiting the recursive least squares (RLS) method under incomplete observation situation. This paper also estimates abnormal flows as outlier sparse flows via sparsity maximization in the under-constrained linear-inverse problem by alternating direction method of multipliers (ADMM). Extensive numerical evaluations show that the proposed algorithm achieves faster convergence per iteration of model approximation, and better volume anomaly detection performance compared with state-of-the-art algorithms

\section{Related Work}
\label{Sec:Related Work}

There is a rich literature in anomaly detection algorithms, and other related fields \cite{Modi_JNCP_2013}. They are for example based on signature profiles, artificial neural networks, fuzzy logic, association rules, support vector machines, genetic algorithms, or hybrid techniques. The efficiency of those algorithms depends on their parameters, their configurations, target anomaly types, or network types.  Hence, this section describes related work for anomaly diagnosis from the viewpoint of time-directional analysis of high-dimensional data. Then, especially addressing efficient analysis means of infinite time-directional traffic data accompanied with noise, high dimension and incomplete observation, the state-of-the-art algorithms for online-based subspace tracking, which is actively studied in the machine learning field, are discussed, which are closely related to our work.

\subsection{Anomaly Detection and Identification}
\label{Sec:AnomalydetectionandIdentification}

The first category of anomaly diagnosis is to detect specific kinds of traffic anomalies including network attacks, network failures, and traffic shifts \cite{Casas_ComputerNetwork_2010}. This category operates on individual and independent time series, and analyzes traffic on a particular network link, particular device readings, or particular packet features. Classical forecasting methods (e.g., ARIMA \cite{Hood_IEEETransRe_1997}, Holt-Winters \cite{Brutlag_LISA_2000}, EWMA\footnote{Exponentially Weighted Moving Average. This is equivalent to ARIMA(0,1,1) model.}, Kalman-Filter \cite{Soule_LSNI_2005}), and outliers analysis methods (e.g., Wavelet \cite{Barford_ACM_IM_2002}, Fourier Transform) are used. 

The second category, which this article mainly addresses, is to diagnose anomalous traffic behaviors from a network-wide perspective. Here, the mechanisms collect coarse-grained SNMP link measurements to detect and isolate volume anomalies in OD flows. This category exploits spatial correlations across the time series of traffic from all the links of a network. Representative works are as follows; the Kalman-filtering approach \cite{Soule_IMC_2005} tracks the evolution of OD flows from SNMP measurements, and identifies anomalies as large prediction errors. The OD flows act as the {\it underlying states} of a network traffic system. The states evolve over time as the OD flows evolve, but the states are not directly observable. Namely, this approach estimates both spatial and temporal correlations \cite{Soule_IMC_2005}. However, it requires a long training period, in which direct anomaly-free OD flow measurements are used to calibrate the model. In addition, these methods do not explicitly handle noisy and higher dimensional characteristics of networks associated with a large number of nodes and links. An efficient lower dimensional approach, i.e., subspace-based approach or low-rank approximation based approach, is desirable to robustly model network traffic behavior. 

A PCA (Principle Component Analysis)-based approach \cite{Lakhina_SIGCOMM_2004, Zhang_ACM_IMC_2005}, one of the subspace-based approaches, separates SNMP measurements into a normal subspace and an anomalous subspace. The success of this approach lies on two points: link traffic is a linear combination of OD flows, and each OD flow has a low intrinsic dimensionality. This approach detects an anomaly when the magnitude of the projection onto the anomal subspace exceeds an associated PCA Q-Statistic threshold \cite{Jackson_Tech_1979}. A critical problem is that it faces scalability problems in large scale networks due to the expensive calculation cost of a big size of its transformation matrix. 
Further subspace-based approaches are proposed in \cite{Roughan_IEEEACM_TranNW_2012, Mardani_IEEEJSTSP_2013, Mardani_IEEEACMTranNW_2015}.  Here, similar to the PCA-based approach, the flow matrix is approximated by a low-rank matrix and sparsity constraints (outlier detection). \cite{Roughan_IEEEACM_TranNW_2012} handles missing data explicitly. A robust PCA-like approach is also proposed to handle anomaly traffic as outlier traffic \cite{Wanga_CN_2012} based on the Relaxed Principal Component Pursuit \cite{Candes_JACM_2011}. Another approach is the SSA-based approach \cite{Tahereh_arXiv_2014}. SSA (singular spectrum analysis) is a technique of time-series analysis, and is a nonparametric spectral estimation method \cite{Broomhead_Physica_1986}. The time-series data is transformed into a Hankel matrix followed by the singular value decomposition to model its time structure. Multivariate SSA (M-SSA) is an extension of basic SSA into multivariate data. Nevertheless, all of these methods consider only a batch-based operation. 

Other than for lower dimensional approaches, which work on a snapshot of the network information, in this paper, we address the difficulty to learn {\it a priori} how such anomalies appear in traffic volume statistics, because large networks are affected by various types of anomalies in different ways. This motivates us to consider an online-based anomaly diagnosis algorithm. An extension of the Holt-Winters forecasting algorithm supports incremental model updating via exponential smoothing \cite{Brutlag_LISA_2000}. A stochastic approximation of the Expectation-Maximization (EM) algorithm for a Gaussian mixture model is proposed in \cite{Hajji_IEEETranNN_2005}.  With respect to the PCA-based approach, although this is a purely spatial algorithm and cannot locate the anomaly temporally, an online formulation of this is proposed that uses a sliding window implementation to identify the normal and abnormal subspaces based on a previous block of time \cite{Lakhina_SIGMETRICS_2004}. Furthermore, a distributed algorithm has been proposed in \cite{Huang_NIPS_2006}. However, since the PCA-based detection algorithm is extremely sensitive to the proper determination of the associated Q-statistics threshold, a straightforward extension into an online-based algorithm is not robust \cite{Ahmed_INFOCOM_2007}. To solve this issue, a kernel version of the recursive least squares algorithm is proposed to construct and adapt a dictionary of features that approximately spans the subspace of normal behavior \cite{Ahmed_INFOCOM_2007}. This uses, however, the number of individual IP flows. A recursive estimation of a flow matrix using a Kalman filtering approach is also proposed \cite{Casas_TC_2009}, but it  uses direct OD flow measurements to calibrate the flow model.

Finally, regarding tensor-based algorithms, a higher-order PCA detection algorithm is proposed based on the Higher-Order singular value decomposition and the Higher Order Orthogonal Iteration \cite{Kim_CAMSAP_2009}. The proposed methods outperform the normal PCA with respect to the scalability of the network size. However, this only considers the direct measurement case, and does not consider network tomography. The evolution of its subspace over time is not also considered. In addition, another anomaly detection algorithm is proposed using the higher order robust PCA with the subspace distance measurement, but this does not also consider network tomography \cite{Zoltowski_GlobalSIP_2014}.

\subsection{General Online-based Subspace Methods for High-dimensional Data Analysis}

This section details general online-based subspace learning methods that our approach falls into. They have been actively studied in machine learning field recently, and can be applied to circumvent potential issues in network analysis especially for noisy, high-dimensional and incomplete measurements.

With regard to matrix-based online algorithms, a representative research is the projection approximation subspace tracking (PAST) \cite{Yang_IEEESP_1995}. GROUSE \cite{Balzano_Conf_2010} recently proposes an incremental gradient descent algorithm performed on the Grassmannian, the set of all $d$-dimensional subspaces of $\mathbb{R}^n$. The algorithm minimizes an $\ell_2$-norm cost function. GRASTA \cite{He_CVPR_2012} enhances robustness against outliers by exploiting an $\ell_1$-norm cost function. pROST proposes an improved GRASTA based on $\ell_0$-surrogates by using the conjugate gradient method \cite{Seidel_MVA_2014}. PETRELS \cite{Chi_IEEETransSP_2013} calculates the underlying subspace via a discounted recursive process for each row of the subspace matrix in parallel. Meanwhile, as for tensor-based online algorithms, an adaptive algorithm to obtain the Candecomp/PARAFAC decompositions \cite{Nion_IEEETransSP_2009} and an accelerated online tensor learning algorithm based on the Tucker decomposition \cite{Yu_ICML_2015} are proposed. However, they do not deal with missing data presence. Online imputation algorithms based on the Candecomp/PARAFAC decomposition are proposed for the presence of missing data  \cite{Kasai_IEEEICASSP_2016_s,Mardani_IEEETransSP_2015}. While \cite{Kasai_IEEEICASSP_2016_s} considers the RLS-based updates, it does not consider time-structured data and anomaly detection, \cite{Mardani_IEEETransSP_2015} considers the stochastic gradient descent (SGD) for large-scale data, and is applied to analyze network anomalies. Nevertheless, its convergence speed of \cite{Mardani_IEEETransSP_2015} is not fast, and the problem definition and the formation of traffic matrices in \cite{Mardani_IEEETransSP_2015} are not the same as ours, thereby it cannot be directly compared with our proposed algorithm.

As seen above, none of these works has provided a complete and reliable solution to model and diagnose high-dimensional large scale data of network traffic data under incomplete observation in an online manner, which is the focus of our approach presented in this paper.

\section{Notations}
\label{sec:Preliminaries}	
Before we present our problem formulation, we summarize the notations used in the remainder of this article. We denote scalars by lower-case letters $(a, b, c, \ldots)$, vectors as bold lower-case letters $(\vec{a}, \vec{b}, \vec{c}, \ldots)$, and matrices as bold-face capitals $(\mat{A}, \mat{B}, \mat{C}, \ldots)$. An element at $(i,j)$ of a matrix \mat{A} is represented as $\mat{A}_{i,j}$. If $\mat{A}$ has additional index like $\mat{A}[t]$ or $\mat{A}$ is a matrix product like $\mat{A}=\mat{BC}$, we use $(\mat{A}[t])_{i,j}$ or $(\mat{BC})_{i,j}$ with parenthesis. $i$-th row vector and $j$-th column of $\mat{A}$ are represented as $\mat{A}_{i,:}$ and $\mat{A}_{:,j}$, respectively. We should particularly note that the transposed column vector of $i$-th row vector $\mat{A}_{i,:}$ is specially denoted as $\vec{a}^i$ in order to explicitly express a row vector, i.e., a horizontal vector. $\mat{A}_{i,p:q}$ represents $(\mat{A}_{i,p}, \ldots, \mat{A}_{i,q}) \in \mathbb{R}^{1 \times (q-p+1)}$. We call a multidimensional or multi-{\it way} (also called {\it order} or {\it mode}) array as {\it tensor}, which is denoted by $(\mathbfcal{A}, \mathbfcal{B}, \mathbfcal{C}, \ldots)$. Similarly, an element at $(i,j,k)$ of a third-order tensor $\mathbfcal{A}$ is expressed as $\mathbfcal{A}_{i,j,k}$.
Tensor {\it slice} matrices are defined as two-dimensional matrices of a tensor, defined by fixing all but two indices. For example, a {\it horizontal slice} and a {\it frontal slices} of a third-order tensor $\mathbfcal{A}$ are denoted as $\mathbfcal{A}_{i,:,:}$ and $\mathbfcal{A}_{:,:,k}$, respectively. Since $\mathbfcal{A}_{:,:,k}$ is heavily used in this article, it is simply expressed as $\mat{A}_k$ using the bold-face capital font and one single subscript in order to explicitly represent its matrix form. Finally, $\vec{a}[t]$ and $\mat{A}[t]$ with the {\it square bracket} represent the computed $\vec{a}$ and $\mat{A}$ after performing $t$-times updates (iterations) in the online-based subspace tracking algorithm described in Section \ref{Sec:ProposedOnlineTrafficTensorAnomography}.
The notation {\rm diag}(\vec{a}), where \vec{a} is a vector, stands for the diagonal matrix with $\{\vec{a}_i\}$ as diagonal elements.
We follow the tensor notation of the review article  \cite{Kolda_SIAMReview_2009} throughout our article and refer to it for further details.

\section{Network Anomography}
\label{Sec:NetworkAnomography}

This section formally defines {\it network tomography} and \ {\it network anomography} \cite{Zhang_ACM_IMC_2005}. For this purpose, we summarize our assumptions. First, we assume that aggregated link measurements (i.e., link matrix) are available via SNMP, and the routing information (i.e., routing matrix) at each time can be obtained from them, such as IGP link weights and the network topology information. Here, we consider the following {\it generative traffic model}. Let $\vec{f}_{i} \in \mathbb{R}^{1\times T}$ be $i$-th traffic for a time period of $T$ between the $i$-th ($i=\{1, \ldots, F\}$) node pair sorted by in a certain order, where $F$ represents the number of flows. This is generated by adding a normal traffic, $\vec{f}_{i}^{(no)} \in \mathbb{R}^{1\times T}$, an anomaly traffic $\vec{f}_{i}^{(ano)} \in \mathbb{R}^{1\times T}$ and a noise $\vec{f}_{i}^{(noise)} \in \mathbb{R}^{1\times T}$ as $\vec{f}_{i}  =  \vec{f}_{i}^{(no)} + \vec{f}_{i}^{(ano)} + \vec{f}_{i}^{(noise)}$. Then, we obtain the flow matrix \mat{F} as $[(\vec{f}_{1})^T : \cdots : (\vec{f}_{i})^T : \cdots :(\vec{f}_{F})^T]^T$ $\in \mathbb{R}^{F \times T}$. Especially, $\mat{F}_{:,t}$, that is $\mat{F}$ at time $t (0 \leq t \leq T)$, is $(\vec{f}_{1}(t), \ldots, \vec{f}_{F}(t))^T \in \mathbb{R}^F$. Here, without losing generality, we assume that routing paths are {\it static} for each pair of nodes during this time period of $T$ because they can be adopted each time in case of a dynamic case. Then, a routing matrix $\mat{R} \subset \{0,1\} \in \mathbb{R}^{L \times F}$ is $\mat{R}_{l,i}=1$ when the flow $\vec{f}_{i}$ passes $l$-th link, $\mat{R}_{l,i}=0$ otherwise, where $L$ represents the number of {\it directly connected} links. Subsequently, the link traffic at time $t$ is represented as  $\mat{Y}_{:,t}  =  \mat{R} \mat{F}_{:,t},$ and, an entire link matrix $\mat{Y} \in \mathbb{R}^{L\times T}$ is represented as $\mat{Y} =  \mat{R}\mat{F}$.

Now we define the problem formulation of network tomography. The relationship among $\mat{F},\ $\mat{R} and $\mat{Y}$ can be reformulated by considering errors as $\mat{Y} = \mat{R} \mat{F} + \mat{E}$, where \mat{E} is the error matrix of size $\mathbb{R}^{L \times T}$. If $\mat{Y}$ is observable measurement and the errors are assumed to be \emph{i.i.d.} Gaussian, the ideal flow matrix $\hat{\mat{F}}$ can be modeled by minimizing the sum-of-squared errors, i.e.,
\begin{equation}
	 \hat{\mat{F}}  =  \defargmin_{\scriptsize{\mat{F}}}  \frac{1}{2}\| \mat{Y} - \mat{R} \mat{F}\|_F^2. \nonumber
\end{equation}
This is an under-constrained or ill-posed inverse problem because the number of OD pairs (unknown quantities), $F$, is more than that of link measurements, $L$, that is $L \ll F$. 

We define network anomography according to \cite{Zhang_ACM_IMC_2005} as follows. Assume that the flow matrix \mat{F} consists of {\it normal flows}, \mat{X}, and {\it abnormal flows}, \mat{V}, as $\mat{F}  =  \mat{X} +  \mat{V}$, where $\mat{X} = [\vec{x}_1: \cdots : \vec{x}_L]$ and $\mat{V} = [\vec{v}_1 : \cdots : \vec{v}_L]$. Then, $\hat{\mat{X}}$ and $\hat{\mat{V}}$ are calculated below;
\begin{equation}
	\label{Eq:ProblemRe-formulation}
		 \{\hat{\mat{X}}, \hat{\mat{V}}\}  =  \defargmin_{\scriptsize{\mat{X}, \mat{V}}}  \frac{1}{2} \| \mat{Y} - \mat{R}(\mat{X}+ \mat{V}) \|_F^2. 
\end{equation}
If \mat{X} is fixed to $\hat{\mat{X}}$, $\hat{\mat{V}}$ is solved by the ill-posed linear inverse problem as 
$ \hat{\mat{V}}   = {\rm arg\ min}_{\scriptsize{\mat{V}}}  \frac{1}{2} \|\mat{D} - \mat{R} \mat{V} \|_F^2$, 
where $\mat{D} = \mat{Y} - \mat{R}\hat{\mat{X}}$ is a model approximation error. Here, we consider how to calculate $\mat{D}$ in existing modeling methods. Time-series analysis techniques (e.g. ARMA, ARIMA, EWMA) derive $\mat{D}$ as forecasting errors. $\mat{D}$ in signal processing based techniques (e.g. Fourier transform, Wavelet transform) is derived from high/middle frequency components by ignoring the lower frequency component. In subspace-based approaches like PCA, $\mat{D}$ is obtained from the abnormal subspace, that is the residual subspace of normal subspace projected by principle components. 

Regarding the abnormal flow $\hat{\mat{V}}$ estimation from signals, a greedy algorithm \cite{Lakhina_SIGCOMM_2004} is proposed, which finds the single largest anomaly in each time instance. Another algorithm is a linear-inverse-based algorithm \cite{Zhang_ACM_IMC_2005}, which calculates the abnormal flow $\hat{\mat{V}}$ by solving the ill-posed linear inverse problem above. This has two types of algorithms. The Frobenius-norm minimization algorithm yields the optimal solution for the Gaussian noise assumption as $\hat{\mat{V}} = \mat{R}^{\dagger} \mat{D} = (\mat{R}^T \mat{R})^{-1} \mat{R}^T (\mat{Y} - \mat{R}\hat{\mat{X}})$,
where $\mat{R}^{\dagger}$ is the pseudo-inverse of \mat{R}. On the other hand,
the $\ell_p$-norm minimization with $0<p \leq 1$, which is called the  {\it sparsity maximization}, considers the errors to be sparsely distributed but possibly large in magnitude. In this case, the $\tau$-th column of $\hat{\mat{V}}$ are the solutions to
\begin{equation}
	\label{Eq:SparsityMaximization}
	\hat{\vec{v}}_{\tau}  =  \defargmin_{\vec{v}_{\tau}}  \| \vec{v}_{\tau} \|_p {\rm \ \ \ s.t.\ \ } \vec{y}_{\tau} -\mat{R}	 \hat{\vec{x}}_{\tau} = \mat{R} \vec{v}_{\tau}.
\end{equation}
This article particularly focuses on the sparsity maximization algorithm for the abnormal flow estimation because this shows superior performances compared to others in \cite{Zhang_ACM_IMC_2005}.

\section{Proposed Online Traffic Tensor Anomography}
\label{Sec:ProposedOnlineTrafficTensorAnomography}

\begin{figure}[t]
	\begin{center}
	\includegraphics[width=14cm, bb=0 0 1417 1151]{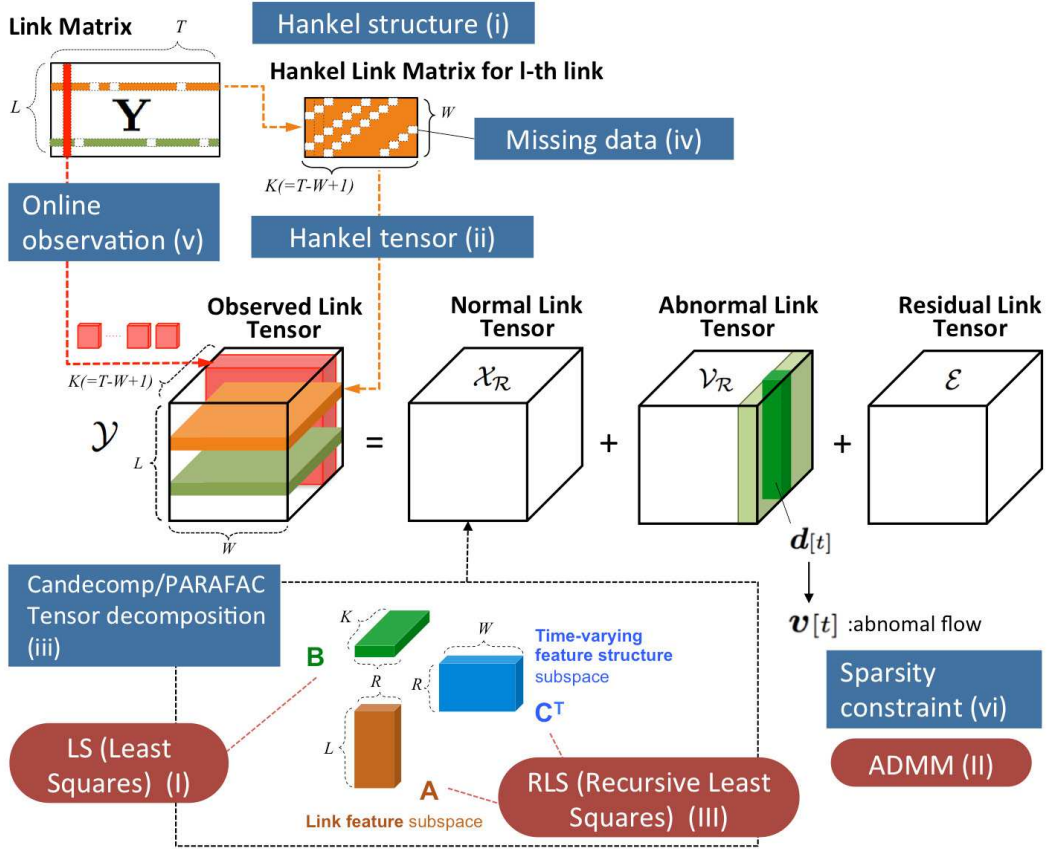}
	\caption{Basic architecture and procedures of the proposed algorithm.}
	\label{Fig:BasicConcept}
	\end{center}
\end{figure}

This section defines the optimization problem of our proposed algorithm, and provides detailed solutions. The overall concept and procedures are summarized in Fig.\ref{Fig:BasicConcept}.

\subsection{Derivation of Problem Formulation}
\label{Sec:DerivationofProblemFormulation}

To robustly model both, the underlying latent structure of normal flows as well as abnormal outlier flows from noisy high-dimensional link measurements, we reformulate the problem (\ref{Eq:ProblemRe-formulation}) by considering its low-rank constraint of \mat{RX} and the sparsity constraint of \mat{V} as 
\begin{equation}
	\label{Eq:Problem-LowRank-Sparsity}
	\min_{\scriptsize{\mat{X}, \mat{V}}} \frac{1}{2} \| \mat{Y} - \mat{R}(\mat{X}+ \mat{V}) \|_F^2 + 
	\mu_r \cdot {\rm rank}(\mat{R}\mat{X}) + \mu_s \| \mat{V}\|_0, 
\end{equation}
where $\mu_r$ and $\mu_s$ control the rank constraint and the sparsity constraint, respectively. By following the literature \cite{Srebro_LearnTheory_2005}, the rank constraint is transformed by the decomposed rank-$R$ matrices $\mat{M}^T \mat{Q}$ of $\mat{RX}$ where $\mat{M} \in \mathbb{R}^{M \times R}$ and $\mat{Q} \in \mathbb{R}^{T \times R}$. Additionally replacing the $\ell_0$-sparsity constraint with its convex $\ell_1$-surrogate, we obtain from (\ref{Eq:Problem-LowRank-Sparsity}) 
\begin{equation}
	\label{Eq:Problem-LowRank-Sparsity-l1}
	\min_{\scriptsize{\mat{X}, \mat{V}}}   \frac{1}{2} \| \mat{Y} - \mat{R}(\mat{X}+\mat{V}) \|_F^2  
	+\mu_r ( \|\mat{M}\|^2_F + \|\mat{Q}\|^2_F ) +\ \mu_s \| \mat{V}\|_1. 
\end{equation}

Now, we address how to capture the time-directional structure of multiple links. Even if we analyze (\ref{Eq:Problem-LowRank-Sparsity-l1}) by keeping the matrix form, we can capture the correlation between multiple links based on a similarity of those temporal variations. However, we cannot deal with the similarity between partial temporal variations inside one single link. In fact, network traffic has periodic and seasonality characteristics accompanied with relatively large noise and fluctuation. This should be taken into account to capture time-spatial correlations among multiple links. Thus, the multidimensional matrix, i.e., tensor, with the {\it Hankel structure} plays a crucial rule by exploiting a three-directional model against noisy and fluctuated signals. See Appendix \ref{Append_Sec:hankel} for the Hankel matrix.  Although this idea is shared with the M-SSA based approach \cite{Tahereh_arXiv_2014}, which combines multiple matrices horizontally, the tensor-based representation of multi-dimensional data can efficiently describe temporal-spatial correlations than M-SSA because the tensor-based approach stacks multiple data into different directions instead of placing them side by side onto the same direction.  

To this end, we first generate the Hankelized time-structured traffic tensor $\mathbfcal{Y}$, which is created by embedding a one-dimensional time-series data into multi-dimensional series. More concretely, let $\{y^l_{1}, \ldots ,y^l_{T}\} \in \mathbb{R}^{1\times T}$ be one-directional time-series traffic volume passing through $l$-th link of length $T$. Given a window length $W$, with $1<W<T$, we construct $k$-th $W$-lagged vectors $\vec{h}^l_{k} = (y^l_{k}, \ldots , y^l_{k+W-1})^T \in \mathbb{R}^W, k = 1,2,\ldots,K$, where $K=T-W+1$, and compose these vectors into the matrix $\mat{H}^l = [\vec{h}^l_{1}:\cdots :\vec{h}^l_{K}] \in \mathbb{R}^{W \times K}$ (Fig.\ref{Fig:BasicConcept}-{\sf i}). By plugging this Hankel matrix $\mat{H}^l$ into the $l$-th horizontal slice matrix of $\mathbfcal{Y}$, that is $\mathbfcal{Y}_{l,:,:}$, the traffic tensor $\mathbfcal{Y}$ is finally generated (Fig.\ref{Fig:BasicConcept}-{\sf ii}). It should be noted that the obtained traffic tensor $\mathbfcal{Y}$ results in $L \times W\times K$ size.

Next, we attempt to model $\mathbfcal{Y}$ as $\mathbfcal{X}_{\mathbfcal{R}}+\mathbfcal{V}_{\mathbfcal{R}}+\mathbfcal{E}$, where $\mathbfcal{X}_{\mathbfcal{R}}$, $\mathbfcal{V}_{\mathbfcal{R}}$, and $\mathbfcal{E}$ are its constituent {\it normal link tensor}, {\it abnormal link tensor} and {\it residual link tensor} with the same size, respectively. 
Here, we model the normal link tensor $\mathbfcal{X}_{\mathbfcal{R}}$ as a low-rank subspace structure from the noisy traffic tensor $\mathbfcal{Y}$ in order to efficiently and robustly capture the change of underlying latent traffic structure of normal flows. For this purpose, particularly addressing the Candecomp/PARAFAC decomposition as a low-rank tensor approximation, we decompose the $\tau$-th frontal slice matrix of $\mathbfcal{X}_{\mathbfcal{R}}$ as $(\mathbfcal{X}_{{\mathbfcal{R}}})_{:,:,\tau} 
 = \mat{A} {\rm diag}(\vec{b}^{\tau}) \mat{C}^T
$, where $\mat{A} = [(\vec{a}^1)^T: \cdots: (\vec{a}^L)^T]^T$, 
$\mat{B} = [(\vec{b}^1)^T: \cdots : (\vec{b}^T)^T]^T$, and
$\mat{C} = [(\vec{c}^1)^T : \cdots : (\vec{c}^W)^T]^T$ with $\{\vec{a}^l, \vec{b}^t,\vec{c}^w\} \in \mathbb{R}^{R}$ (Fig.\ref{Fig:BasicConcept}-{\sf iii}). See Appendix \ref{Append_Sec:CPDEC} for a brief introduction of the Candecomp/PARAFAC decomposition. Subsequently, we obtain the transformed problem formula of (\ref{Eq:Problem-LowRank-Sparsity-l1}) by additionally considering the normal link tensor $\mathbfcal{X}_{\mathbfcal{R}}$ with the Hankel structure as well as the Candecomp/PARAFAC tensor decomposition as
\begin{equation}  
\label{Eq:HankelTensorProblemFormulation}
\begin{split}
\min_{\scriptsize{\mat{A},\mat{B},\mat{C},\mat{V}}} \ \ &
\frac{1}{2} \| 
\underbrace{\mathbfcal{P}_{{\Omega}}}_{\rm Missing\ data}
\underbrace{\bigl[{\mathbfcal{Y}} -(\mathbfcal{X}_{\mathbfcal{R}}+\mathbfcal{V}_{\mathbfcal{R}}) \bigr]}_{\rm Approximation\ error} \|_F^2  + 
\underbrace{\mu_r (\| \mat{A}\|_F^2 + \| \mat{B}\|_F^2 + \| \mat{C}\|_F^2)}_{\rm Frobenius-norm\ regularizer}
+\!\! \underbrace{\mu_s \|\mat{V}\|_1,}_{\rm Sparsity\ regularizer}\\
 {\rm s.t.}\   & \underbrace{(\mathbfcal{X}_{{\mathbfcal{R}}})_{:,:,\tau} 
 = \mat{A} {\rm diag}(\vec{b}^{\tau}) \mat{C}^T}_{\rm Candecomp/PARAFAC\ constraint}, {\rm \ and}
\!\! \underbrace{(\mathbfcal{X}_{{\mathbfcal{R}}})_{l,:,:} \in \mathcal{S}_H,}_{\rm Hankel\ structure\ constraint}
\end{split}
\end{equation}
where $\mathbfcal{Y} = \mathbfcal{H}(\mat{Y})$, $\mathbfcal{X}_{\mathbfcal{R}} = \mathbfcal{H}(\mat{RX})$, and $\mathbfcal{V}_{\mathbfcal{R}}  = \mathbfcal{H}(\mat{RV})$, and $\mathbfcal{H}(\cdot)$ is the operator of the Hankelization of the matrix into the tensor structure. $\mathcal{S}_H$ represents the constraint of the linear subspace of all matrices with the Hankel structure. It should be noted that the Hankel structure constraint is placed on only $\mathbfcal{X}_{\mathbfcal{R}}$ in this paper for simplicity. In addition, the Frobenius-norm regularization offers a viable option for a batch-based low-rank tensor decomposition under the Candecomp/PARAFAC model \cite{Bazerque_IEEETransSP_2013}. Moreover, taking into account {\it incomplete observation} situation, the formulation in (\ref{Eq:HankelTensorProblemFormulation}) additionally considers the support of the analysis with interpolating missing measurements (Fig.\ref{Fig:BasicConcept}-{\sf iv}), which is generally called the {\it tensor completion} problem. For this purpose, $\mathbfcal{P}_{{\Omega}}(\cdot)$ represents the operator to extract observation data, more precisely, $\mathbfcal{P}_{\Omega}(\mathbfcal{X})_{i_1,i_2,i_3}=\mathbfcal{X}_{i_1,i_2,i_3}$ if $(i_1, i_2, i_3) \in \Omega$ and $\mathbfcal{P}_{\Omega}(\mathbfcal{X})_{i_1,i_2,i_3}=0$ otherwise. $\Omega$ is a subset of the complete set of indices $\{(i_1, i_2, i_3): i_d \in \{1, \ldots, n_d \}, d \in \{1,2,3\}$.

Finally, we consider an {\it online-based} setting of subspace learning, outlier learning, and anomaly detection method to prevent all measurements and model parameters in the past from being stored (Fig.\ref{Fig:BasicConcept}-{\sf v}). To this end, we  tackle an {\it online tensor completion} problem. It should be noted that, as for the Hankel structure constraint of $\mathcal{S}_H$, this present paper considers {\it only} the two successive slice matrices to avoid model re-construction and diagonal-averaging using $\{\vec{b}^\tau\}_{\tau=t-W+1}^{\tau=t-1}$ in the past. Consequently, instead of (\ref{Eq:HankelTensorProblemFormulation}), the final problem of our proposed method is further formulated to estimate the Candecomp/PARAFAC factor matrices $\{\mat{A}, \vec{b}, \mat{C}\}$ and the abnormal flow matrix $\mat{V}$ by considering the exponential weighted least squares cost function  
\begin{equation}  
\label{Eq:Final_Problem_Definition}
\begin{split}
        \min_{\scriptsize{\mat{A},\vec{b},\mat{C},\mat{V}}}  \displaystyle{\frac{1}{2}}& \underbrace{\sum_{\tau=1}^t \lambda^{t-\tau}}_{\rm Online-based} \!
        \biggl[ 
        \underbrace{{\Big\| \overbrace{{\rm\bf P}_{{\bf \Omega}_{\tau}}}^{\rm Missing\ data}  \Bigl( \mat{Y}_{\tau} - \left(\mat{A} {\rm diag}(\vec{b}^\tau) \mat{C}^T 
        \!+\! {\mat{V}_{{\scriptsize \mat{R}}_{\tau}}} \right) \Bigr) \Big\|_F^2}}_{\rm Approximation\ error\ by\ low-rank\ Candecomp/PARAFAC}
        \\
    +&\ \mu_h [\tau] \Big\| \underbrace{{\rm\bf P}_{({\bf \Omega}_{\tau})_{:,1:W\!-\!1}}}_{\rm Missing\ data} \underbrace{\Bigl( (\mat{A} {\rm diag}(\vec{b}^{\tau\!-\!1}) \mat{C}^T)_{:,2:W}  
    - \ (\mat{A} {\rm diag}(\vec{b}^\tau) \mat{C}^T)_{:,1:W\!-\!1} \Bigr) \Big\|_F^2}_{\rm Hankel\ structure\ error}  \\   
    +& \underbrace{\bar{\mu}_r[\tau](\| \mat{A}\|_F^2 + \| \mat{C}\|_F^2) + \mu_r [\tau] \| \vec{b}^\tau \|_2^2}_{\scriptsize \rm Frobenius\ and\ \ell_2\ norm\ regularizer\ for\ \{\mat{A},\vec{b},\mat{C}\}}\ + \!\!\!\!\!\! \underbrace{\mu_s[\tau] \| \vec{v}[\tau] \|_1}_{\scriptsize \rm Sparsity\ regularizer\ for\ \mat{V}}    \!\!\!\!\!\!
        \biggr],
\end{split}
\end{equation}   
where ${\rm\bf P}_{{\bf \Omega}_\tau}(\cdot)$ is the matrix linear operator of the $\tau$-th frontal slice matrix of $\mathbfcal{P}_{{\Omega}}(\cdot)$. $\mat{Y}_\tau$ and $\mat{V}_{{\scriptsize \mat{R}_\tau}}$ are the $\tau$-th frontal slice matrices of  $\mathbfcal{Y}$ and $\mathbfcal{V}_{\mathbfcal{R}}$, respectively. $\vec{v}[\tau]$ is the $W$-th column of $\mat{V}_{{\scriptsize \mat{R}_\tau}}$. $\bar{\mu}_r[\tau]=\mu_r[\tau]/\sum_{j=1}^\tau\lambda^{\tau-j}$, and we choose a lower $\mu_r$ when the data can be assumed to have a lower error rate, a relatively higher value otherwise. $0 < \lambda \leq 1$ is the so-termed {\it forgetting factor}. When $\lambda < 1$, data in the past are exponentially down-weighted, which facilitates tracking in non-stationary environments. In the case of infinite memory $\lambda = 1$, this coincides with the batch-based estimator.

We finally perform the anomaly detection and identification. A larger value in $\vec{v}$ indicates that the probability of the existence of anomaly in its flow is higher. Therefore, by introducing a threshold variable $\delta_v$, the flows of which $\vec{v}$ is larger than $\delta_v$ are categorized into abnormal flows at time $t$.

\subsection{Solutions and Algorithm}
\label{Sec:DetailedSolutionsandAlgorithm}

This section gives the detailed solutions of the final minimization problem (\ref{Eq:Final_Problem_Definition}), where unknown variables are $\vec{v}, \mat{A}, \mat{C}$, and $\vec{b}$. It is readily seen that this function is not convex. However, if $\mat{A}$, $\mat{C}$ and $\vec{b}$ are fixed, the problem becomes convex in  $\vec{v}$. Similarly, if $\vec{v}$ fixed, we can  refine our estimate of \mat{A}, \mat{C} and $\vec{b}$ in successive convex optimization steps. This paper uses an alternating minimization procedure to successively solve lower-dimensional convex problems by updating the unknown variables alternatively.

\subsubsection{Update of \vec{b}$[t]$ by LS (Fig.\ref{Fig:BasicConcept}: I)}
\label{Sec:UpdateOfB}

We calculate $\vec{b}[t]$ via an $\ell_2$-norm regularized least squares (LS) problem, which has a closed-form solution. The estimate $\vec{b}[t]$ is obtained by calculating (\ref{Eq:Final_Problem_Definition}) with fixed $\{ \mat{A}[t-1], \mat{C}[t-1] \}$ derived at time $t\!-\!1$, i.e.,
\begin{equation}
\begin{split}
\label{eq:Problem_Definition_b_timebase}
    \vec{b}[t] =  \defargmin_{\vec{b} \in \mathbb{R}^R} &\frac{1}{2} 
    \Biggl[ \|  {\bf \Omega}_t \circledast 
    [ \mat{Y}_t -  \mat{A}[t\!-\!1] {\rm diag}(\vec{b}) (\mat{C}[t\!-\!1])^T ] \|_F^2 
    \\
    & 
    + \mu_h[t] \| ({\bf \Omega}_t)_{:,1:W\!-\!1} \circledast \bigl[ (\mat{A}[t\!-\!1] {\rm diag}(\vec{b}[t\!-\!1]) \mat{C}[t\!-\!1]^T)_{:,2:W}  \\
    &- (\mat{A}[t\!-\!1] {\rm diag}(\vec{b}) \mat{C}[t\!-\!1]^T)_{:,1:W\!-\!1} \bigr] \|_F^2
    + \mu_r[t] \| \vec{b} \|_2^2 \Biggr], \nonumber \\
\end{split}
\end{equation}
where ${\bf \Omega}_t$ denotes a $L\times W$ binary $\{0,1\}$-matrix with $({\bf \Omega}_t)_{l,w}=1$ if 
$\mathbfcal{Y}_{l,w,t}$ is observed, and $({\bf \Omega}_t)_{l,w}=0$ otherwise.
$\circledast$ represents the Hadamard product, i.e., the element-wise product of matrices. Defining $F[t]$ as the inner objective to be minimized, we obtain $\vec{b}[t]$ since $\vec{b}[t]$ satisfies $\partial F[t]/\partial \vec{b}[t] = 0$ as
\begin{align}
\label{Eq:b_final}
\vec{b}[t] = & 
\biggl[ 
\mu_r[t] \mat{I}_R +  
\sum_{l=1}^L  \sum_{w=1}^W ({\bf \Omega}_t)_{l,w} 
 \vec{g}_{l,w}[t]  (\vec{g}_{l,w}[t])^T + 
 \mu_h[t] \sum_{l=1}^{L}  \sum_{w=1}^{W-1}  ({\bf \Omega}_t)_{l,w}  (\vec{g}_{l,w}[t]) (\vec{g}_{l,w}[t])^T \biggr]^{-1} 
\nonumber \\
& \biggl[
\sum_{l=1}^L \sum_{w=1}^W 
({\bf \Omega}_t)_{l,w}  (\mat{Y}[t])_{l,w} 
\vec{g}_{l,w}[t]
+
\mu_h[t] \sum_{l=1}^{L} \sum_{w=1}^{W-1} ({\bf \Omega}_t)_{l,w}  (\vec{g}_{l,w+1}[t])^T  \vec{b}[t\!-\!1] \vec{g}_{l,w}[t]
\biggr],
\end{align} 
where $\vec{g}_{l,w}[t] = \vec{a}^l[t-1] \circledast \vec{c}^w[t-1] \in \mathbb{R}^R$.

\subsubsection{Update of $\vec{v}[t]$ based on ADMM (Fig.\ref{Fig:BasicConcept}: II)}      
\label{Sec:UpdateOfV}

$\vec{v}[t]$ is solved by the {\it alternating direction method of multipliers} (ADMM), which solves convex optimization problems by separating them into smaller sub-problems \cite{Boyd_FTML_2011}. It has recently gained big attention in wide applications in a number of areas. Thus, $\vec{v}[t]$ is obtained by solving the reformulated  (\ref{Eq:Final_Problem_Definition}) by addressing only the last column of each frontal slice matrix  as
\begin{equation}
\begin{split}
	\label{eq:}
        \vec{v}[t]& = \defargmin_{\scriptsize{\vec{v}}}  {\frac{1}{2} 
        \| {\bf \Omega}_t \circledast \Bigl( \mat{Y}_t }  -  \mat{A}[t\!-\!1] {\rm diag}(\vec{b}[t]) (\mat{C}[t\!-\!1])^T- {\mat{V}_{\scriptsize \mat{R}}}_t \Bigr)
         \|_F^2 +\ \mu_s[t] \| \vec{v} \|_1 \nonumber\\
        & =  \defargmin_{\scriptsize{\vec{v}}}  \frac{1}{2} 
        \|  \vec{q}[t] - \mat{R}_{\omega_{t, W}}\vec{v}  \|_2^2  +\ \mu_s[\tau] \| \vec{v} \|_1,
\end{split}
\end{equation}
where $\vec{q}[t] \!=\! ({\bf \Omega}_t)_{:,W} \!\circledast\! \bigl[ (\mat{Y}_t)_{:,W} \!-\!(\mat{A}[t\!-\!1] {\rm diag}(\vec{b}[t]) (\mat{C}[t\!-\!1])^T)_{:,W}\bigr] \in \mathbb{R}^L$, and $({\bf \Omega}_t)_{:,W}$ has only the last column of ${\bf \Omega}_t$. Furthermore, the routing matrix $\mat{R}_{\omega_{t, W}} \in \mathbb{R}^{L \times F}$, which corresponds to only observed measurements, is calculated as 
$\mat{R}_{\omega_{t, W}}={\rm diag}(({\bf \Omega}_t )_{:,W})\mat{R}$. Thereby, this problem can be re-written as $\min f(\vec{v}) + g(\vec{z}) \ {\rm s.t.} \ \vec{v} - \vec{z} = 0$, where $f(\vec{v}) = (1/2) \| \vec{q}[t] - \mat{R}_{\omega_{t, W}}\vec{v} \|_2^2$, and $g(\vec{z}) = \mu_s[t] \| \vec{z} \|_1$. It should be also noted that the iteration index {\it t} is the {\it outer} loop index, and this is kept fixed at the ADMM loop, i.e., the {\it inner} loop, where a new index {\it k} is used instead. The augmented Lagrangian of this constrained minimization problem is expressed as $\mathcal{L}_{\xi}(\vec{v}, \vec{z}, \vec{y}) =  f(\vec{v}) + g(\vec{z}) + \vec{y}^T(\vec{v} -\vec{z}) + (\xi/2) \| \vec{v} -\vec{z} \|^2_2$, where \vec{y} is the dual vector. Denoting $\vec{u} = (1/\xi) \vec{y}$ as the {\it scaled dual variable}, the sub-problems of ADMM become
\begin{subnumcases}
{}
    \label{Eq:v_final_a}
    \vec{v}^{k+1} =  ( \mat{R}_{\omega_{t, W}}^T \mat{R}_{\omega_{t, W}} + \xi \mat{I}_{F})^{-1} (\mat{R}_{\omega_{t, W}}^T \vec{q}[t] + \xi (\vec{z}^{k} -\vec{u}^{k} )) \ \ \ \ & \\
   \label{Eq:v_final_b}
    \vec{z}^{k+1}  =  S_{\mu_s[t]/\xi} (\vec{v}^{k+1} + \vec{u}^{k+1}) & \\
    \label{Eq:v_final_c}
    \vec{u}^{k+1} =  \vec{u}^{k}  + \vec{v}^{k+1} - \vec{z}^{k+1},&
\end{subnumcases}
where $\xi > 0$ and $S_{\kappa}(a)$ is the {\it soft thresholding operator} that is defined as $S_{\kappa}(a) = {\rm sign}(a)(|a|-\kappa)_{+}$, namely, $S_{\kappa}(a)=0$ if $|a|\leq \kappa$, otherwise $S_{\kappa}(a)={\rm sign}(a)(|a|-\kappa)$. Finally, we obtain $\vec{v}^{k+1}$ as $\vec{v}[t]$. 
The overall algorithm for $\vec{v}[t]$ is summarized in {\bf Algorithm \ref{Alg:Admm}}.
\begin{algorithm}
\caption{Calculate the abnormal flow vector $\vec{v}$ at $t$}
\label{Alg:Admm}
\begin{algorithmic}[1]
\REQUIRE{The absolute tolerance $\epsilon^{abs}$, the absolute relative $\epsilon^{rel}$, ADMM maximum iteration $K$.}
\STATE{Initialize $\vec{v}_{1}=\vec{v}^{0}, \vec{z}_{1}=\vec{z}^{0}, \vec{y}_{1}=\vec{y}^{0}$. \\
(either to zero or to the final value from the last subspace update of the same data vector for a warm start.)}
\FOR{$k=1,2, \ldots, K$} 
\STATE{Update the abnormal flow vector \vec{v}.
		\hfill (\ref{Eq:v_final_a})
			 }
\STATE{Update the dual vector \vec{z}.
                \hfill (\ref{Eq:v_final_b})
}
\STATE{Update the scale parameter \vec{u}.
               \hfill (\ref{Eq:v_final_c})
}
\STATE{Calculate primal and dual residuals $r^{pri}$ and $r^{dual}$.}
\STATE{Update stopping criteria $\epsilon^{pri}$ and $\epsilon^{dual}$ using $\epsilon^{abs}$ and $\epsilon^{rel}$.}
\IF{$r^{pri} \leq \epsilon^{pri}$ and $r^{dual} \leq \epsilon^{dual}$} \STATE{Converge and break the loop.} \ENDIF
\ENDFOR
\RETURN $\vec{v}[t]=\vec{v}^{k+1}, \vec{z}[t]=\vec{z}^{k+1}, \vec{y}[t]=\vec{y}^{k+1}$.
\end{algorithmic}
\end{algorithm}

\subsubsection{Update of \mat{A}$[t]$ and \mat{C}$[t]$ by RLS (Fig.\ref{Fig:BasicConcept}: III)}
\label{Sec:UpdateOfAandC}

The calculation of $\mat{C}[t]$ requires $\mat{A}[t\!-\!1]$, and the calculation of $\mat{A}[t]$ uses $\mat{C}[t]$. This paper addresses a second-order stochastic gradient based on the RLS method with forgetting parameters, which has been widely used in tracking of time varying parameters in many fields. Its computation is efficient since we update the estimates recursively every time new data becomes available. First, the problem (\ref{Eq:Final_Problem_Definition}) is reformulated to obtain $\mat{A}[t]$ as
\begin{equation}
\begin{split}
\label{eq:Problem_Definition_A}
\min_{\scriptsize \mat{A}}& \frac{1}{2} \sum_{\tau=1}^t \lambda^{t-\tau} 
\biggl[ \| {\bf \Omega}_\tau \circledast \bigl[ \mat{Y}_\tau 
 \!\!-\!\!  (\mat{A} {\rm diag}(\vec{b}[\tau]) (\mat{C}[\tau\!\!-\!\!1])^T \!\!-\!\! {\mat{V}_{\scriptsize \mat{R}}}_\tau
) \bigr] \|_F^2   \\
&+ \| ({\bf \Omega}_\tau)_{:,1:W\!-\!1} \circledast \bigl[ (\mat{A} {\rm diag}(\vec{b}[\tau\!-\!1]) \mat{C}[\tau\!\!-\!\!1]^T)_{:,2:W} \\
&- (\mat{A} {\rm diag}(\vec{b}[\tau]) (\mat{C}[\tau\!\!-\!\!1])^T)_{:,1:W\!-\!1} \bigr] \|_F^2
\biggr]+ 
\frac{\mu_r[t]}{2} \| \mat{A}\|_F^2.
\end{split}
\end{equation}
The objective function in (\ref{eq:Problem_Definition_A}) is decomposed into a parallel set of smaller problems, one for each row $\vec{a}^l \in \mathbb{R}^R$ of $\mat{A}$. By denoting $\mat{Y}_\tau-{\mat{V}_{\scriptsize \mat{R}}}_\tau$ as $\mat{Z}_\tau$, we obtain $\vec{a}^l[t]$ as
\begin{equation}  
\begin{split}
	\label{eq:problem_def_am}
	\min_{\vec{a}^l \in \mathbb{R}^R} & \frac{1}{2}  
	\sum_{\tau=1}^t 
	\Biggl[
	\sum_{w=1}^W
	\lambda^{t-\tau} \!\!
	\left( ({\bf \Omega}_{\tau})_{l,w} \!
	\left((\mat{Z}_\tau)_{l,w} 
	\!\!-\!(\vec{a}^l)^T {\rm diag} (\vec{b}[\tau]) \vec{c}^w[\tau\!\!-\!\!1]\right) \right) ^2  \\ 
	&+\ \mu_h \sum_{w=1}^{W\!-\!1} \lambda^{t-\tau} 
	\left(({\bf \Omega}_{\tau})_{l,w}  \left(
	(\vec{a}^l)^T {\rm diag} (\vec{b}[\tau\!\!-\!\!1]) \vec{c}^{w+1}[\tau\!\!-\!\!1] 
	- 
	(\vec{a}^l)^T {\rm diag} (\vec{b}[\tau]) \vec{c}^w[\tau\!\!-\!\!1]\right)\right)^2
	\Biggr] \nonumber \\
	&+\ \frac{\mu_r[t]}{2}\| \vec{a}^l\|_2^2.
	\end{split}
\end{equation}
\begin{algorithm}[t]
\caption{Online algorithm for subspace tracking and anomaly detection}
\label{Alg:overall}
\begin{algorithmic}[1]
\REQUIRE{ $\{ \mat{Y}[t]$ and ${\bf {\Omega}}[t] \}^{\infty}_{t=1}$, $\lambda$, $\mu_r$, $\mu_h$}
\STATE{Initialize \{$\mat{A}[0]$, $\vec{b}[0]$, $\mat{C}[0]$\} and \mat{Y}[0]=\mat{0}.}
\FOR{$t=1,2, \ldots$} 
\STATE{Update the projection coefficient vector $\vec{b}[t]$. \hfill (\ref{Eq:b_final})}
\STATE{Calculate the abnormal flow vector $\vec{v}[t]$ via {\bf Algorithm 1} using ADMM.
}
\STATE{Detect abnormal flows from $\vec{v}[t]$.}	
\STATE{Update subspace factor matrices $\mat{C}[t]$. \hfill (\ref{Eq:cw_final})}		
\STATE{Update subspace factor matrices $\mat{A}[t]$. \hfill (\ref{Eq:al_final})}
\ENDFOR
\end{algorithmic}
\end{algorithm}
Here, denoting ${\rm diag}(\vec{b}[\tau])\vec{c}^w[\tau\!-\!1]$ and ${\rm diag}(\vec{b}[\tau\!-\!1]) \vec{c}^{w+1}[\tau\!-\!1]-{\rm diag}(\vec{b}[\tau]) \vec{c}^w[\tau\!-\!1]$ as $\vec{\alpha}_w[\tau] \in \mathbb{R}^R$ and $\vec{\beta}_w[\tau] \in \mathbb{R}^R$, respectively, $\vec{a}^l[t]$ is obtained by setting the derivative of (\ref{eq:problem_def_am}) with regard to $\vec{a}^l$ equal to zero. 
\begin{eqnarray}
	\label{Eq:al_final}
	\vec{a}^l[t] 
	&=& \vec{a}^l[t\!-\!1] \!-\! (\mat{RA}_l[t])^{-1} \Bigl(
	\mu_h\sum_{w=1}^{W\!-\!1} ({\bf \Omega}_t)_{l,w+1} \vec{\beta}_w[t]  (\vec{\beta}_w[t])^T 
	+ (\mu_r[t] - \lambda \mu_r[t\!-\!1]) \mat{I}_R \Bigr) 
	 \vec{a}^l[t\!-\!1] \nonumber \\
	&&+\ (\mat{RA}_l[t])^{-1} \sum_{w=1}^W ({\bf \Omega}_t)_{l,w} \left((\mat{Y}_t)_{l,w} - ({\mat{V}_{\scriptsize \mat{R}}}_t)_{l,w} - (\vec{\alpha}_w[t])^T  \vec{a}^l[t\!-\!1] \right) \vec{\alpha}_w[t], 
\end{eqnarray}
where $\mat{RA}_l[t]$ is calculated as 
\begin{equation}  
\begin{array}{lll}
	\label{eq:Update_RA}
	\mat{RA}_l[t]  
 	&=&  \lambda \mat{RA}_l[t\!-\!1]
	+  \displaystyle{\sum_{w=1}^W ({\bf \Omega}_t)_{l,w+1} \vec{\alpha}_w[t]  \vec{\alpha}_w[t]^T}  \\
	&&+\ \displaystyle{\mu_h  \sum_{w=1}^{W\!-\!1} ({\bf \Omega}_t)_{l,w} \vec{\beta}_w[t] \vec{\beta}_w[t]^T
	+ (\mu_r[t] \!-\! \lambda \mu_r[t\!-\!1] ) \mat{I}_R}.
\end{array}
\end{equation} 
The derivations of (\ref{Eq:al_final}) and (\ref{eq:Update_RA}) are detailed in Appendix \ref{Append_Sec:RLS}.
Meanwhile, as for $\mat{C}[t]$, the problem (\ref{Eq:Final_Problem_Definition}) is reformulated as
\begin{equation}  
\begin{split}
	\label{Eq:problem_def_cw}
	\min_{\vec{c}^w \in \mathbb{R}^R} &  \frac{1}{2} 
	\sum_{\tau=1}^t 
	\Biggl[
	\sum_{l=1}^L
	\lambda^{t-\tau} \!\! \left(({\bf \Omega}_{\tau})_{l,w} 
	\!\left((\mat{Z}_\tau)_{l,w} 
	\!\!-\!\! (\vec{a}^l[\tau])^T {\rm diag} 
	(\vec{b}[\tau]) \vec{c}^w \right) \right) ^2  \\ 
	&+\ 	\mu_h \sum_{l=1}^{L} \lambda^{t-\tau} 
	\left(({\bf \Omega}_{\tau})_{l,w}  \left(
	(\vec{a}^l[\tau])^T  {\rm diag} (\vec{b}[\tau\!\!-\!\!1]) \vec{c}^{w+1}[\tau] - 
	(\vec{a}^l[\tau])^T  {\rm diag} (\vec{b}[\tau]) \vec{c}^w \right)\!
	\right)^2
	\Biggr] \\
	&+ \frac{\mu_r[t]}{2}\| \vec{c}^w\|_2^2.\nonumber
\end{split}
\end{equation} 
It should be emphasized that the second term representing the Hankel structure error is not included when $w=W$. In addition,  since the second term needs $\vec{c}^{w+1}[\tau]$, this calculation cannot be performed in parallel, and the order of the calculations follows the {\it descending order} of $w$. Finally, $\vec{c}^w[t]$ can be given by denoting $(\vec{a}^l[\tau])^T {\rm diag}(\vec{b}[{\tau}])$ and $(\vec{a}^l[\tau])^T {\rm diag}(\vec{b}[\tau\!-\!1])$ as $\vec{\gamma}_{l}[\tau] \in \mathbb{R}^{1\times R}$ and $\vec{\eta}_{l}[\tau] \in \mathbb{R}^{1\times R}$, respectively, as 
\begin{eqnarray}
	\label{Eq:cw_final}
	\vec{c}^w[t] 
	 &=& \vec{c}^w[t\!-\!1] - (\mat{RC}_w[t])^{-1} (\mu_r[t]  - \lambda \mu_r[t-1]) \vec{c}^w[t\!-\!1] \nonumber \\
	 & &+\ (\mat{RC}_w[t])^{-1} 
	 \sum_{l=1}^L({\bf \Omega}_t)_{l,w}
	\biggl(
	\Bigl((\mat{Y}_t)_{l,w} - ({\mat{V}_{\scriptsize \mat{R}}}_t)_{l,w}  \bigr. \Biggr. \nonumber\\
	& &+\ \biggl. \Bigl. \mu_h \vec{\eta}_{l}[t] \vec{c}^{w+1}[t] \Bigr) \mat{I}_{R}
	-  (1+\mu_h)\vec{c}^w[t\!-\!1] \vec{\gamma}_{l} [t] 
	\biggr) (\vec{\gamma}_{l}[t])^T, 
\end{eqnarray}
where $\mat{RC}_w[t]$ is transformed as 
\begin{align}
	\label{eq:Update_RC}
	\mat{RC}_w[t]  	& = 
 	\lambda \mat{RC}_w[t-1] 
	+\sum_{l=1}^L  
	({\bf \Omega}_t)_{l,w}(1+\mu_h)\vec{\gamma}_{l} [t] (\vec{\gamma}_{l} [t])^T + (\mu_r[t] - \lambda \mu_r[t\!-\!1]) \mat{I}_R.
\end{align}

The overall algorithm to solve (\ref{Eq:Final_Problem_Definition}) is finally summarized in {\bf Algorithm \ref{Alg:overall}}.

\subsection{Computational Complexity Analysis}
\label{Sec:ComputationalComplexityAnalysis}
This section analyzes the computational complexity per iteration of the proposed algorithm. The calculation of $\vec{b}[t]$ in Section \ref{Sec:UpdateOfB} requires $\mathcal{O}(|{\bf \Omega}_t| R^2)$ in (\ref{Eq:b_final}), where $|{\bf \Omega}_t|$ is the number of known entries in ${\bf \Omega}_t$. The calculations of $\mat{A}[t]$ and $\mat{C}[t]$ in Section \ref{Sec:UpdateOfAandC} require $\mathcal{O}(L R^3)$ for (\ref{Eq:al_final}) and $\mathcal{O}(W R^3)$ for (\ref{Eq:cw_final}), respectively, for the inversion of $\mat{RA}$ in (\ref{eq:Update_RA}) and $\mat{RC}$ in (\ref{eq:Update_RC}). As for the calculation of $\vec{v}[t]$ in \ref{Sec:UpdateOfV}, the inversion of $(\mat{R}_{\omega_{t, W}}^T \mat{R}_{\omega_{t, W}} + \xi \mat{I}_{F})$ can be done efficiently by exploiting the {\it matrix inversion lemma}, which states that  $(\mat{P}+\xi \mat{A}^T \mat{A})^{-1} = \mat{P}^{-1} \mat{A}^T(\mat{I} + \xi \mat{A} \mat{P}^{-1} \mat{A}^T)^{-1} \mat{A}\mat{P}^{-1} $. This leads to $\mathcal{O}(|({\bf \Omega}_t)_{:,W}|F^2)$, where $|({\bf \Omega}_t)_{:,W}|$ is the number of known values at the $t$-th iteration. Then, the total calculation in \ref{Sec:UpdateOfV} needs at most $\mathcal{O}((K+1)|({\bf \Omega}_t)_{:,W}|F^2)$ due to this one-time inversion and $K$-times multiplications for all inner iterations in (\ref{Eq:v_final_a}), where $K$ is the maximum number of inner iterations in {\bf Algorithm 1}. 
Thus, the total computational complexity at $t$-th iteration in {\bf Algorithm 2} results in $\mathcal{O}(|{\bf \Omega}_t| R^2+(L+W)R^3+(K+1)|({\bf \Omega}_t)_{:,W}|F^2))$, and reveals that the number of flows, $F$, is dominant since rank $R$ is assumed to be low-rank.

\section{Numerical Evaluation}
\label{Sec:NumericalEvaluations}

We show numerical comparisons of the proposed algorithm with state-of-the-art algorithms for synthetic and real-world datasets. All the following experiments are done on a PC with 3.0 GHz Intel Core i7 CPU and 16 GB RAM. The synthetic anomalies are injected onto the synthetic and real-world datasets to be evaluated. More concretely, as for the real-world dataset, the original signal in the dataset is firstly smoothed, and the synthetic anomalies are injected onto it.  Regarding the synthetic dataset, the synthetic anomalies are injected onto the synthetic signal that includes the seasonal signal, the periodic signals and the noise. We use the Receiver Operating Characteristic (ROC) and the F-measure value as the evaluation metrics. The ROC evaluates a binary classifier, which plots {\it true positive rate} against {\it false positive rate} at various {\it discrimination thresholds}. In this case, this thresholds correspond to $\delta_v$ in Section \ref{Sec:DerivationofProblemFormulation}. F-measure effectively references the true positives to the arithmetic mean of the predicted positives and the real positives, which is calculated as the harmonic mean between precision and recall. As for the comparison algorithms, the EWMA algorithm and the Wavelet-based algorithm proposed in \cite{Zhang_ACM_IMC_2005} are compared with the proposed algorithm. They provide a unified frame for integrating anomaly detection approaches and the inference techniques for anomaly identifications. The former is one representative method of the time-series modeling, and the latter is one of the signal processing based methods. But the latter is a batch-based method. It should be noted that the standard methods of these do not handle {\it missing data}. Therefore, the latest observed data corresponding a missing datum is interpolated to perform fair comparison. In addition, three subspace-based tracking algorithms, which are GROUSE, GRASTA, and PETRELS, are compared. We use Matlab codes provided by the respective authors.
It is important to note that since these algorithms do not have the abnormal flow detection function, we have newly integrated the function defined in (\ref{Eq:SparsityMaximization}) onto them. For the proposed algorithm, we use $\lambda=0.9, \mu_r=\mu_h=10^{-3}$ and 
$\mu_s=10^{-2}\times \max({\rm abs}(\mat{R}_{\omega_{t, W}}\vec{v}))$ for synthetic datasets, 
and $\mu_s=10^{-1}\times \max({\rm abs}(\mat{R}_{\omega_{t, W}}\vec{v}))$ for real-world datasets, 
where ${\rm abs}(\vec{a})$ returns the absolute value of each element in \vec{a}. $K$, $\epsilon^{abs}$, and $\epsilon^{rel}$ for ADMM are $120$, $10^{-5}$, and $10^{-3}$, respectively.

\subsection{Evaluation Methodology and Synthetic Anomaly Injection}
\label{Sec:SyntheticAnomalyInjection}

The evaluation of any anomaly detection algorithms always faces the issue how to obtain {\it ground truth} because no public and reliable ground truth is available in real-world datates \cite{Soule_IMC_2005}. One popular way is that a security expert labels anomalies by manual inspection against collected live traffic traces. However, this is very expensive and time-consuming if datasets are large. In addition, this is not a perfect solution because the operator could make mistakes that miss an anomaly or generate a false positive anomaly. Furthermore, because such traces include a limited number of anomalies, comprehensive performance evaluations, which evaluate all algorithm capabilities, are difficult to perform. Instead, an alternative approach is to {\it inject synthetic anomalies}, which correspond to the ground truth, onto data signals. One advantage of this is to be able to change the {\it anomaly parameters} assuming various anomalies. In other words, the flexible configurability of anomaly parameters as explained below allows us to simulate a wide range of anomalies that cannot be found in real-world datasets, and this shall enable us to achieve comprehensive evaluations of the anomaly detection and identification performance of algorithms. 
\begin{table}[htbp]
	\begin{center}
	\caption{Synthetic anomaly parameters.}
	\label{Tbl:SyntheticAnomalyInjectionCofig}
	\begin{tabular}{p{4cm}|p{7cm}|p{4cm}}
	\hline
	Parameters & Settings & Description\\
	\hline
	\hline
	Multiplicative ratio of volume change (${\it \delta}$) & $1.5 \leq \delta \leq 2.5$ & DDoS, alpha event\\	
	\cline{2-3}
	   & ${\it \delta} = 0$ & Outrage\\	
	\hline
	Duration (${\it d}$)  & ${\it d} = \{5,10,20,30\}$ (mins) &  \\
	\hline
	Increase ratio (${\it \gamma_{i}}$)  & Change time ratio against entire duration (${\it d}$): $0 \leq {\it \gamma}< 0.5$& Gradual or sudden up/down changes\\		
	\cline{1-1}
	Decrease ratio (${\it \gamma_{d}}$)  &   &  \\		
	\hline
	\hline
	Target OD flows  (Number of flows) & N-1: One flow between one source (src) and one destination (dst) &  DDoS, alpha\\
	\cline{2-3}
	& N-1: N flows between srcs/dess and one src/des & DDoS\\
	\cline{2-3}	
	& All-ODs-one-link: All ODs passing one link  & Outrage\\	
	\hline
	\end{tabular}
	\end{center}
\end{table}
\begin{figure}[t]
	\begin{center}
	\includegraphics[width=\linewidth, bb=0 0 1436 793]{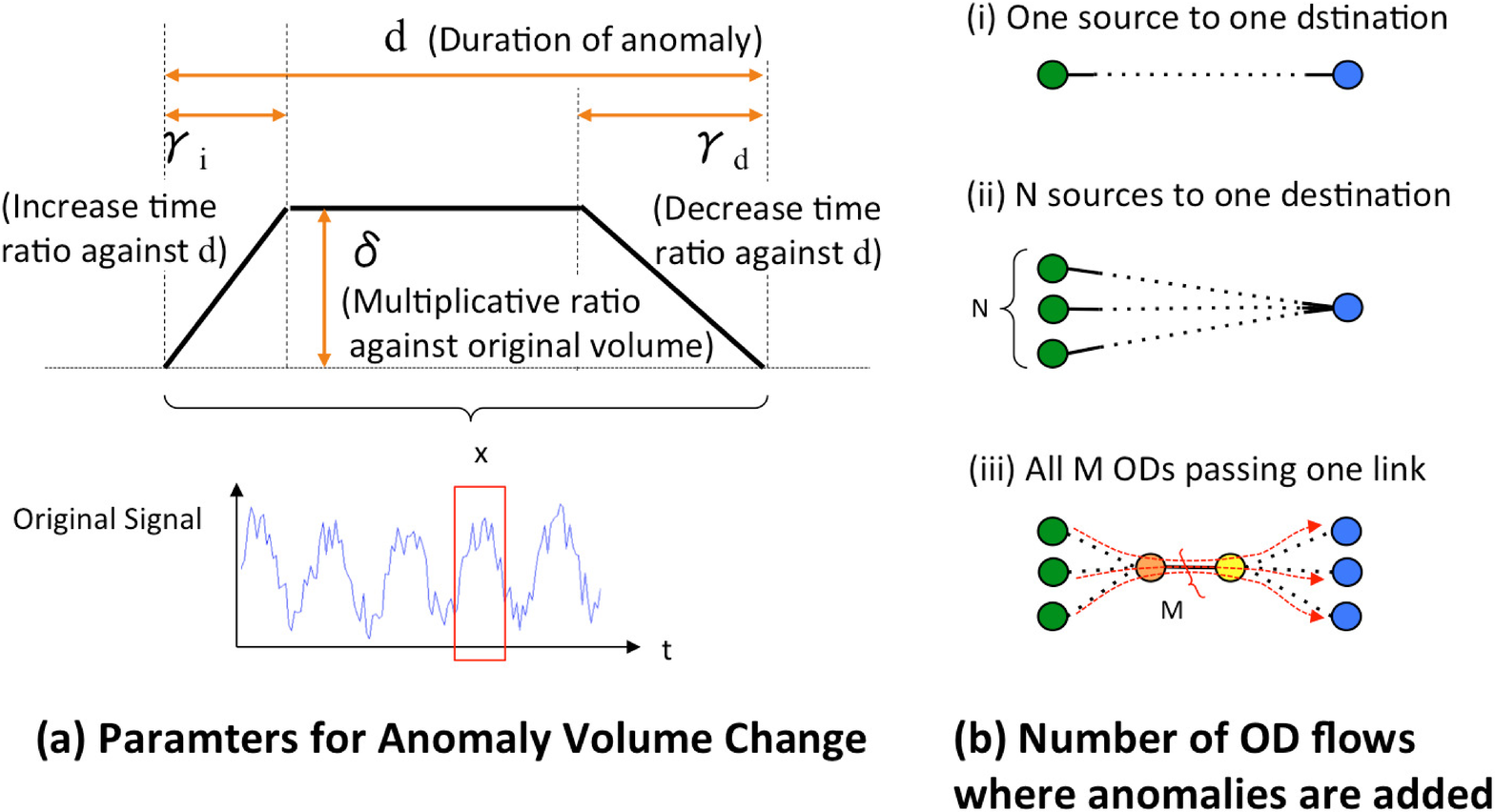}
	\caption{Illustration of the anomaly parameters.}
	\label{Fig:SyntheticAnomalyInjectionCofig}
	\end{center}
\end{figure}

Hereafter, the anomaly parameters are detailed by referring the conceptual illustration shown in Fig.\ref{Fig:SyntheticAnomalyInjectionCofig}, and the summarization in Table \ref{Tbl:SyntheticAnomalyInjectionCofig}. It should be noted that this paper focuses on anomalies that bring {\it changes of volume patterns}, e.g. DDoS attacks, alpha events, outrages, and does not assume worms and port scans.
As for the {\it duration of an anomaly} $d$, most DDoS attacks are observed to continue between $5$ and $30$ minutes, and some outliers last less than $1$ minute and others last several days \cite{Moore_USS_2001}. The DDoS attacks in the Abilene Network dataset last less than $20$ minutes, and some outliers continue more than $2$ hours. The alpha events could be of any length. However, because we cannot simulate all outliers, this paper configures the duration $d$ as $\{5,10,20,30\}$ minutes. 
We mimic the {\it traffic volume change}, when anomalies occur, by introducing a multiplicative factor $\delta$ that is multiplied by the original traffic of a target OD flow. Adding $20\%$, for example, of the original OD flow volume is simulated by using $\delta=1.2$. $\delta=0$ is used to capture the outage scenarios. We can mimic a variety of either the DDoS attacks or the alpha events by allowing $1.0 \leq \delta \leq 2.0$. We do not consider $\delta > 2.5$ because such changes are clearly irregular. We also address the {\it traffic shape} at the beginning and ending of anomalies. The initial rise of the DDoS attacks could be simulated by a ramp shape. The outage anomalies show an almost sudden drop in volume like a square shape. The alpha events indicate either an almost sudden rise or a ramp-shape increase. Thus, we introduce the increase ratio parameter ($\gamma_{i}$) and the decrease ratio parameter ($\gamma_{d}$) to express these shapes. For both parameters, we use $0 \leq \gamma< 0.5$ against the entire duration $d$. Finally, we address the {\it flow structure}, i.e., {\it combination of the number of sources and destinations}, that has an influence on the structure of the OD flows in an entire network. We denote ``1-1" as the OD flow that traverses from one single source to one single destination. This could occur with the DDoS attacks or the alpha events. ``N-1" refers to the OD flows between N-sources and one single destination, which could happen with the DDoS attacks. ``All-ODs-one-link" corresponds to all the OD flows that pass one particular single link.

\subsection{Real-world Dataset Evaluations}

We use the Abilene Network Dataset for the evaluation of the proposed method on real data. Abilene Network is the Internet2 backbone network in the US. It has 11 Points of Presence, where there are $121$ OD flows and $30$ links. The Abilene Network Dataset samples 2016 samples per week, and 5-minutes sampled traffic matrices are collected via Netflow. Each element of a generated flow matrix corresponds to a single OD flow over time with 5 minute increments.
	
The procedure to generate anomaly-injected real-world dataset is explained {by following Section \ref{Sec:SyntheticAnomalyInjection}. An example of the overall procedure is depicted in Fig.\ref{fig:AnomalyInjectionAbilene}, where the case of the $50$th flow of the Abilene Network Dataset is illustrated. The long-term statistical {\it trend} from an original OD flow (Fig.\ref{fig:AnomalyInjectionAbilene}: 1st graph) is extracted by {\it smoothing} the original signal (2nd graph). This is achieved by approximating the extracted signal by 5-th Wavelet levels by Daubechies-5 mother wavelet with $5$ levels because these underlying trends are generally non-stationary. Next, a Gaussian noise with zero mean (3rd graph) is added onto the smoothed, i.e., de-noised signal, where the distribution variance is calculated using the first 5 detailed signals (4th graph). Finally, injecting one of the anomalies (5th graph) onto smoothed noisy signal, we obtain the final anomaly-injected noisy signal (6th graph). 
\begin{figure}[htbp]
	\begin{center}
	\includegraphics[width=\linewidth, bb=0 0 1533 879]{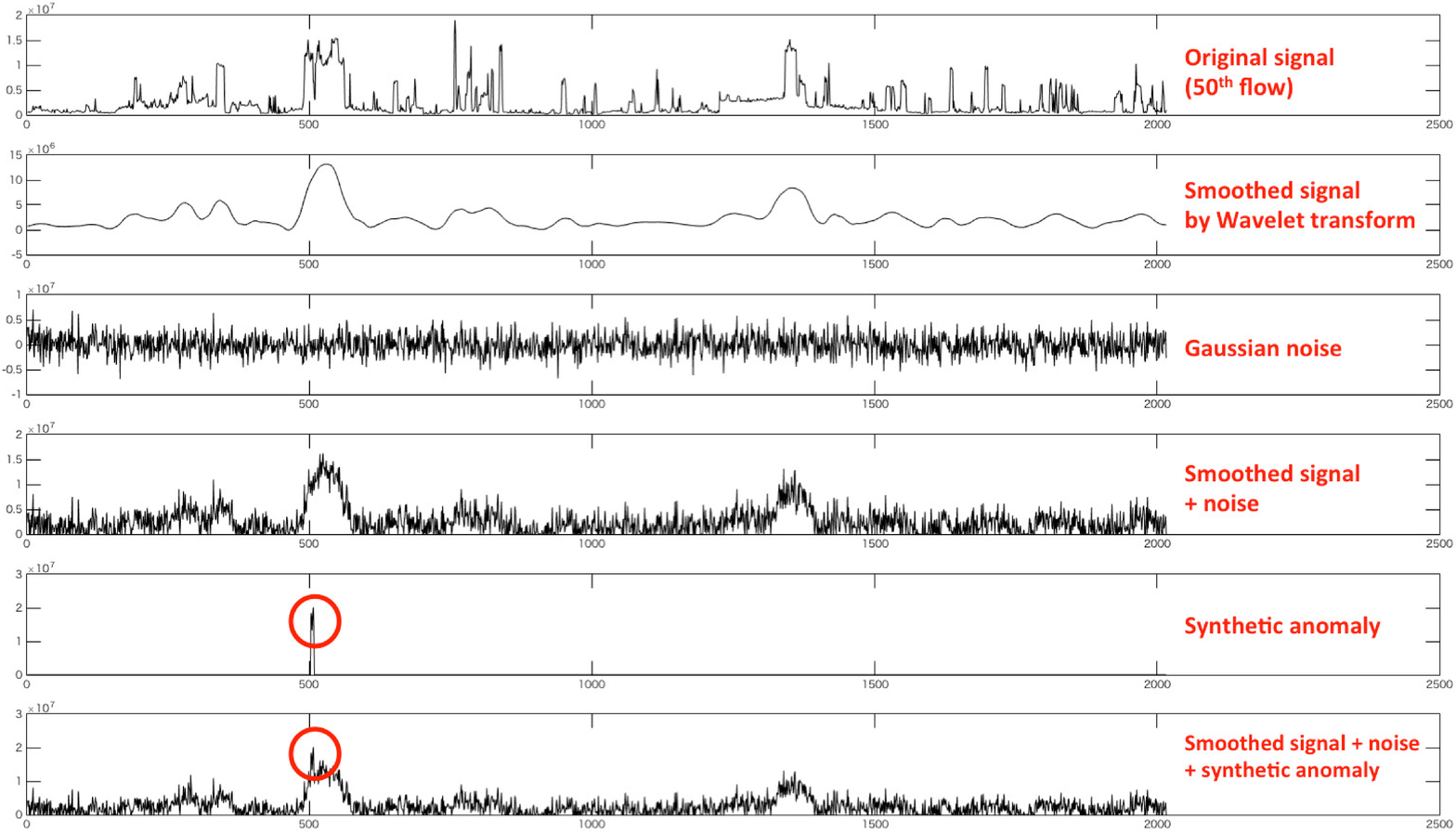}
	
	\caption{Anomaly injection procedure in Abilene (50th flow).}
	\label{fig:AnomalyInjectionAbilene}
	\end{center}	
\end{figure}

We evaluate the case where the sampling frequency is every $5$ minutes, and observation ratio $\rho$ is $30$. Since the number of nodes, $N_{node}$, is $11$, $L=30$ and $F=12$. $W$ is $288$, which corresponds to $1$ day, and $T=2016$ (= $1$ week). The ratio of the anomaly-injected flows is $1.54$\%. The results of the modeling residual error and the ROC curve are shown in Fig.\ref{fig:Real_OR_30}. From the residual error in Fig.\ref{fig:Real_OR_30}(a), the error of Wavelet is much lower than others because, in the Wavelet algorithm, the signal with higher frequency signal is removed and the residual data is not produced by its constructed model. In addition, although the EWMA algorithm constructs a parametrized model, the number of model parameters are much more than that of the subspace-based algorithms. Among the subspace-based algorithms, the proposed algorithm shows the lowest errors. The convergence speed of PETREL is faster than that of GROUSE and GRASTA because PETRELS has a second-order convergence property. On the other hand, the proposed algorithm shows much faster convergence characteristics than that of PETRELS. As for the ROC carve in Fig.\ref{fig:Real_OR_30}(b), the proposed algorithm also outperforms the other algorithms. 
\begin{figure}[htbp]
        \begin{center}
        \includegraphics[width=0.5\hsize]{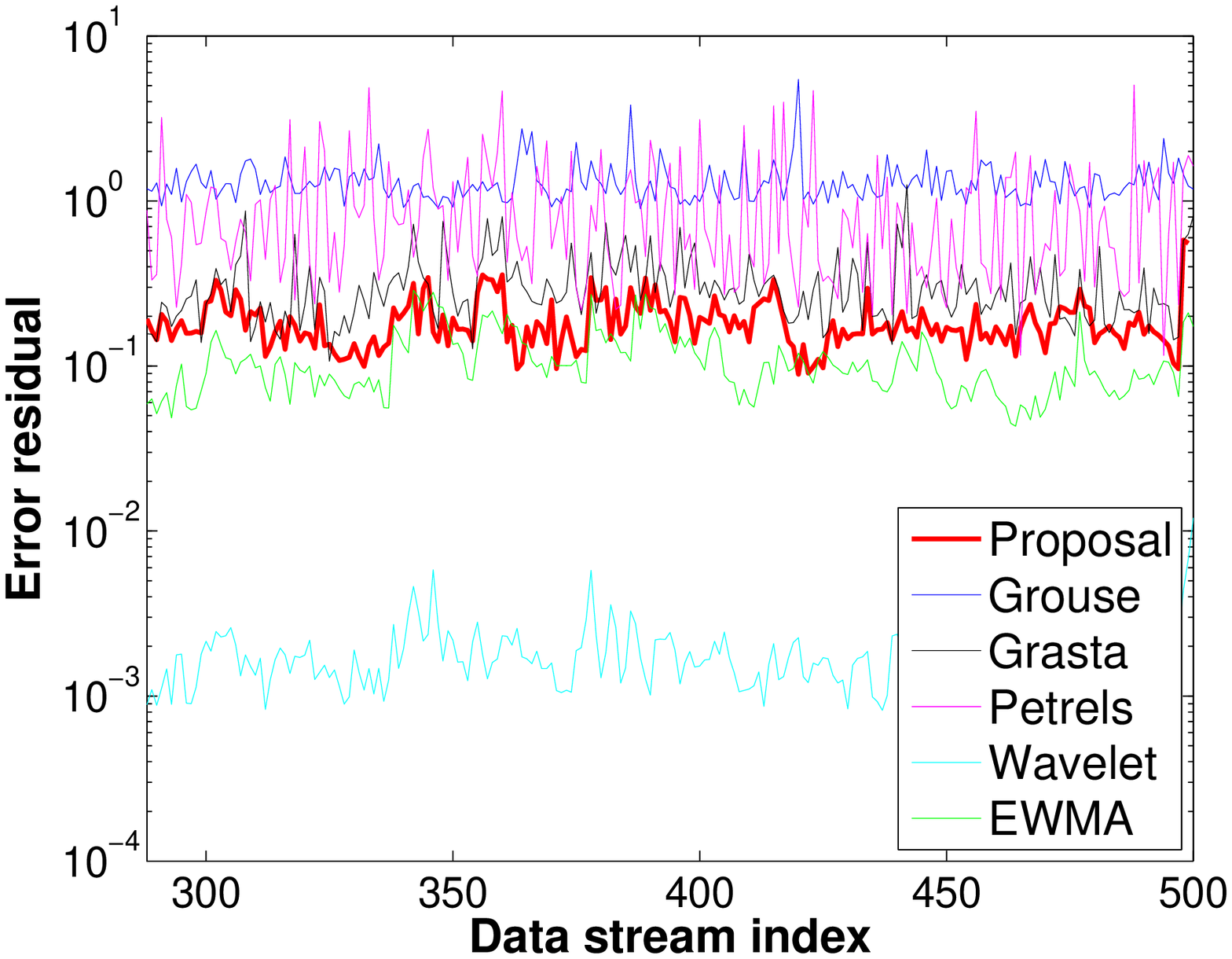}
        
         {(a) Residual error}
        \label{}
        \vspace*{0.3cm}

        \includegraphics[width=0.5\hsize]{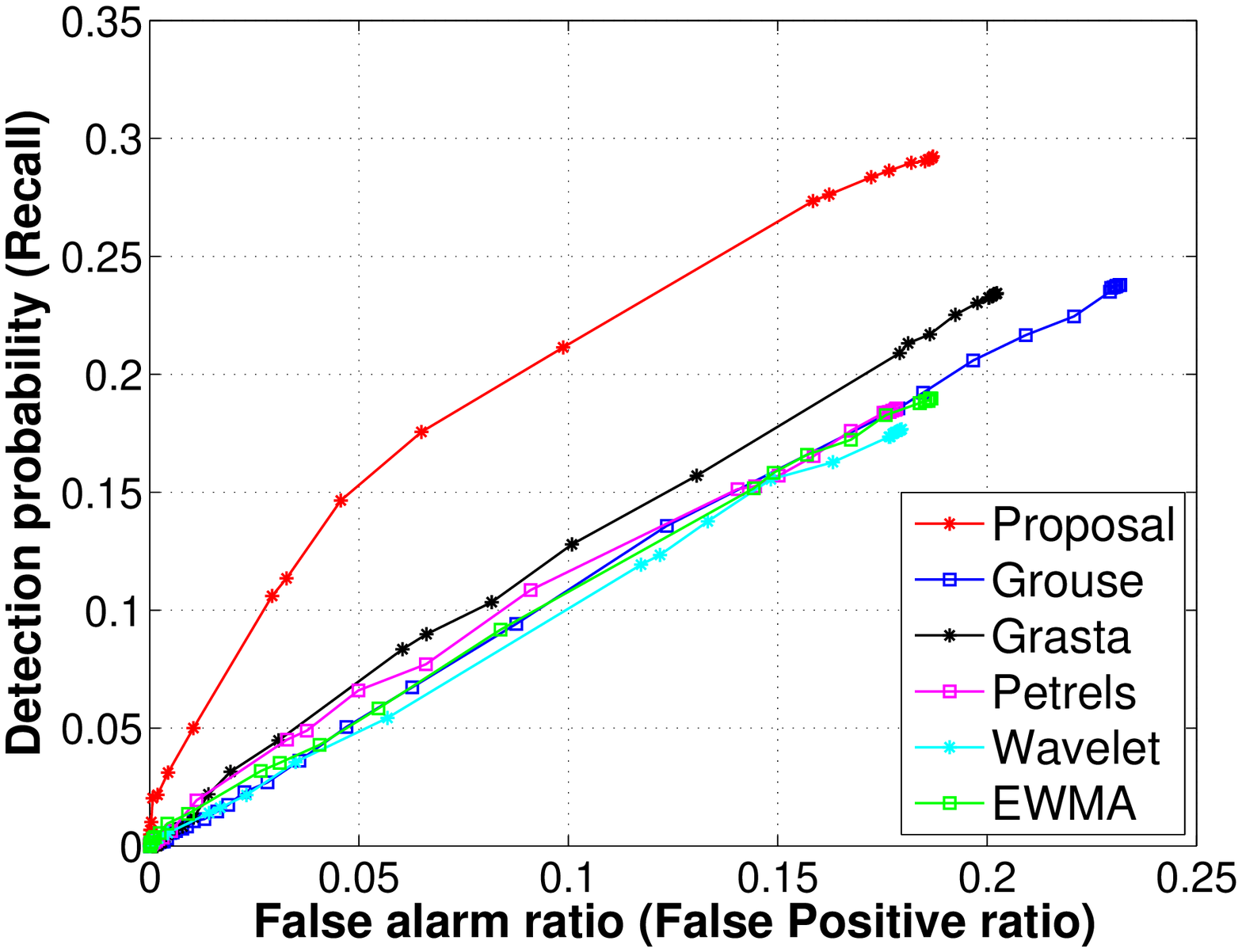}
        
         {(b) ROC}
        \label{}
        \end{center}

	\vspace*{-0.5cm}
    \caption{Residual error and ROC for real-world dataset ($\rho=30$).}
    \label{fig:Real_OR_30}
    \vspace*{0.5cm}
     
        \begin{center}
        \includegraphics[width=0.5\hsize]{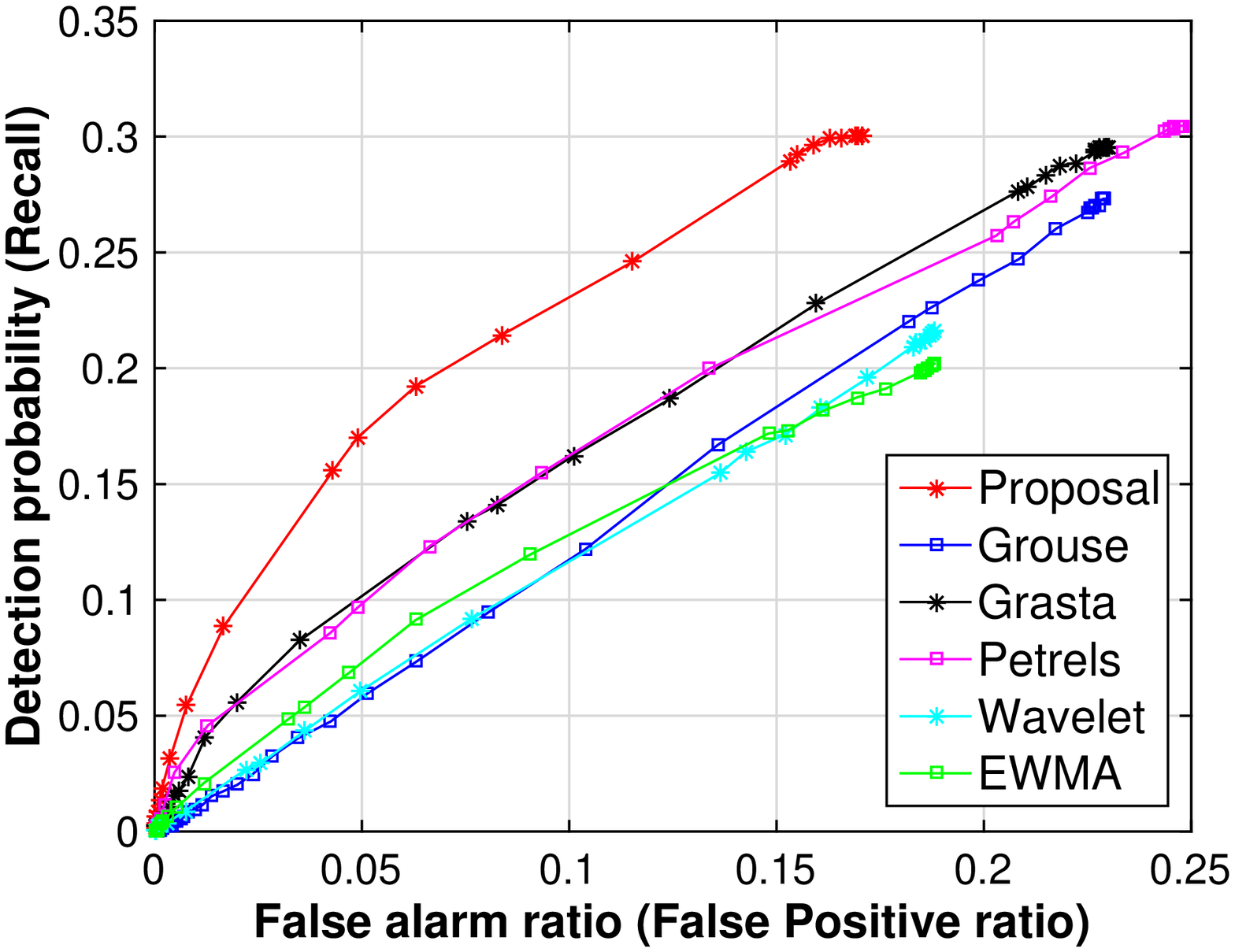}

	\vspace*{-0.2cm}        
	\caption{ROC for real-world dataset ($\rho=50$).}        
        \end{center}
	\label{fig:Real_OR_50}
\end{figure}

The result of the ROC for an  observation ratio $\rho$ of $50$ is also shown in Fig.\ref{fig:Real_OR_50}. This also shows the superior performance of the proposed algorithm against the other algorithms.

\subsection{Synthetic Datasets Evaluations}

As mentioned earlier, we use the synthetic datasets for comprehensive evaluations.

\subsubsection{Network Generation}
\label{sec:Network Generation}

The network used as an input for the simulation is generated in a random fashion. A number of nodes $N_{node}$ are randomly placed on a 2D space, and directly-connecting links among those nodes are calculated by the {\it Delaunay triangulation} algorithm, which maximizes the minimum angle of the triangles. Then, $F$ traffic flows are generated between $F$ pairs of nodes that are randomly selected. Each traffic flow travels its shortest path between its source and destination nodes, which is calculated by the {\it Dijkstra} algorithm.

\subsubsection{Anomaly-injected Traffic Generation}

We generate a virtual flow at time $t$ for the $i$-th node pair by {\it injecting} synthetic anomalies as $\vec{f}_{i}(t)=\vec{f}_{i}^{(po)}(t) + \vec{f}_{i}^{(sea)}(t) + \vec{f}_{i}^{(ano)}(t) + \vec{f}_{i}^{(noise)}(t)$ 
by following Section \ref{Sec:NetworkAnomography} \cite{Kim_CAMSAP_2009}. It is noted that $\vec{f}_{i}^{(no)}(t)$ in Section \ref{Sec:NetworkAnomography} is further decomposed into the {\it periodic component} $\vec{f}_{i}^{(po)}(t)$ and the {\it seasonal trend component} $\vec{f}_{i}^{(sea)}(t)$, where  $\vec{f}_{i}^{(po)}(t)$ is generated as 	$\vec{f}_{i}^{(po)}(t)=A_1 \sin(\omega t)$, and three types of signals are mixed equally for $\vec{f}_{i}^{(sea)}(t)$ as 
\begin{eqnarray}
	\vec{f}_{i}^{(sea)}(t)   =  \left\{
	\begin{array}{lr}
		0 & {\rm (no\ trend)},\\
		b_1 t & {\rm (linear\ trend)},\\
		A_2/b_2 \sin(7\omega t) & {\rm (weekly\ sine\ wave\ trend)}.\nonumber
	\end{array}
	\right.
\end{eqnarray}
$\vec{f}_{i}^{(noise)}(t)$ is created as $\vec{f}_{i}^{(noise)}(t) \sim \mathcal{N}(0, \sigma^2)$, which is a Gaussian noise with a zero-mean and $\sigma^2$ variance. Lastly, we inject the synthetic anomalies $\vec{f}_{i}^{(ano)}(t)$ onto the above $\vec{f}_{i}(t)$, i.e., $\mat{F}_{:,t}$ as explained in Section \ref{Sec:SyntheticAnomalyInjection}. The final link matrix for $L$ link-pairs at each time $t$ is calculated  as $\mat{Y}_{:,t} = \mat{R} \mat{F}_{:,t}$.

\subsubsection{Experimental Results in Small-size Network}

We first consider a small-size network with $N_{node}=500$, where $L$ is $2958$, and $F$ is $5 \times 10^4$. $W$ is $24$, which corresponds to $1$ day,  and $T$ is $168$, i.e., $1$ week. The data is sampled every hour, and the observation ratio $\rho$ is $30$. The ratio of injected anomalies is $1.81\times 10^{-2}$\%. The results are shown in Fig.\ref{fig:Synthetic_SmallSize_OR_30}.
\begin{figure}[htbp]
        \begin{center}
          \includegraphics[width=0.5\hsize]{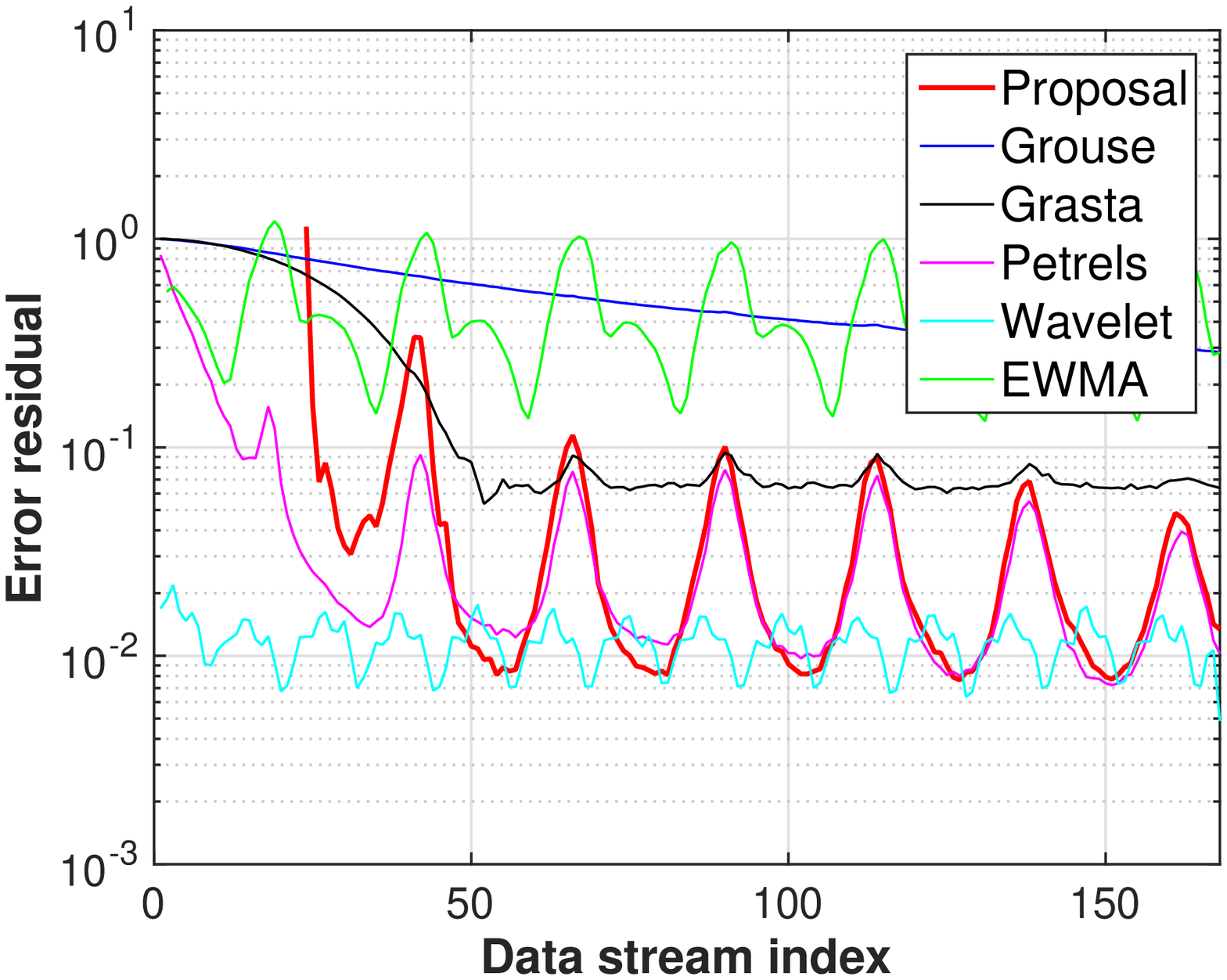}
        
         {(a) Residual error}
        \vspace*{0.4cm}

        \includegraphics[width=0.5\hsize]{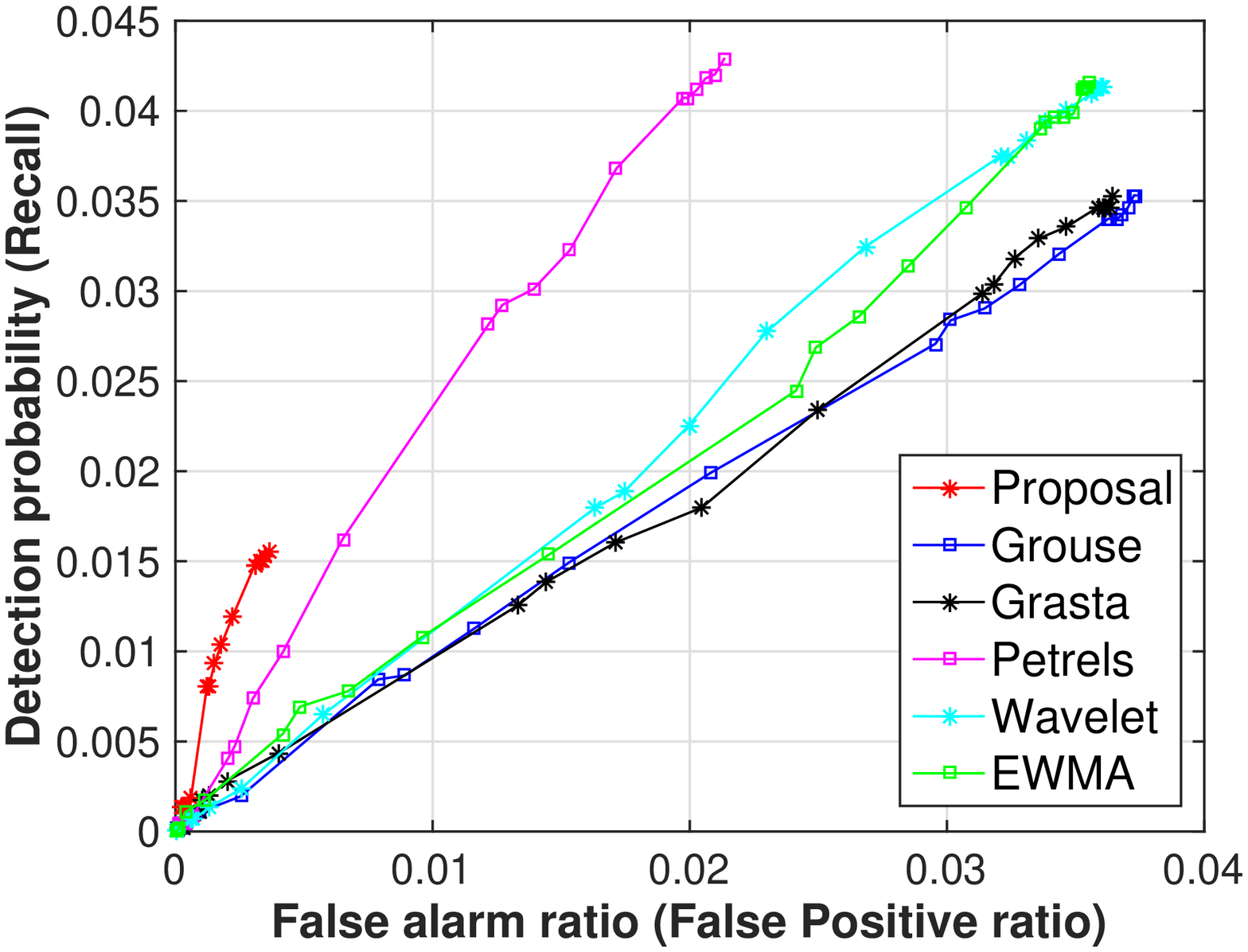}
        
         {(b) ROC}
        \vspace*{0.2cm}        
        
        \includegraphics[width=0.5\hsize]{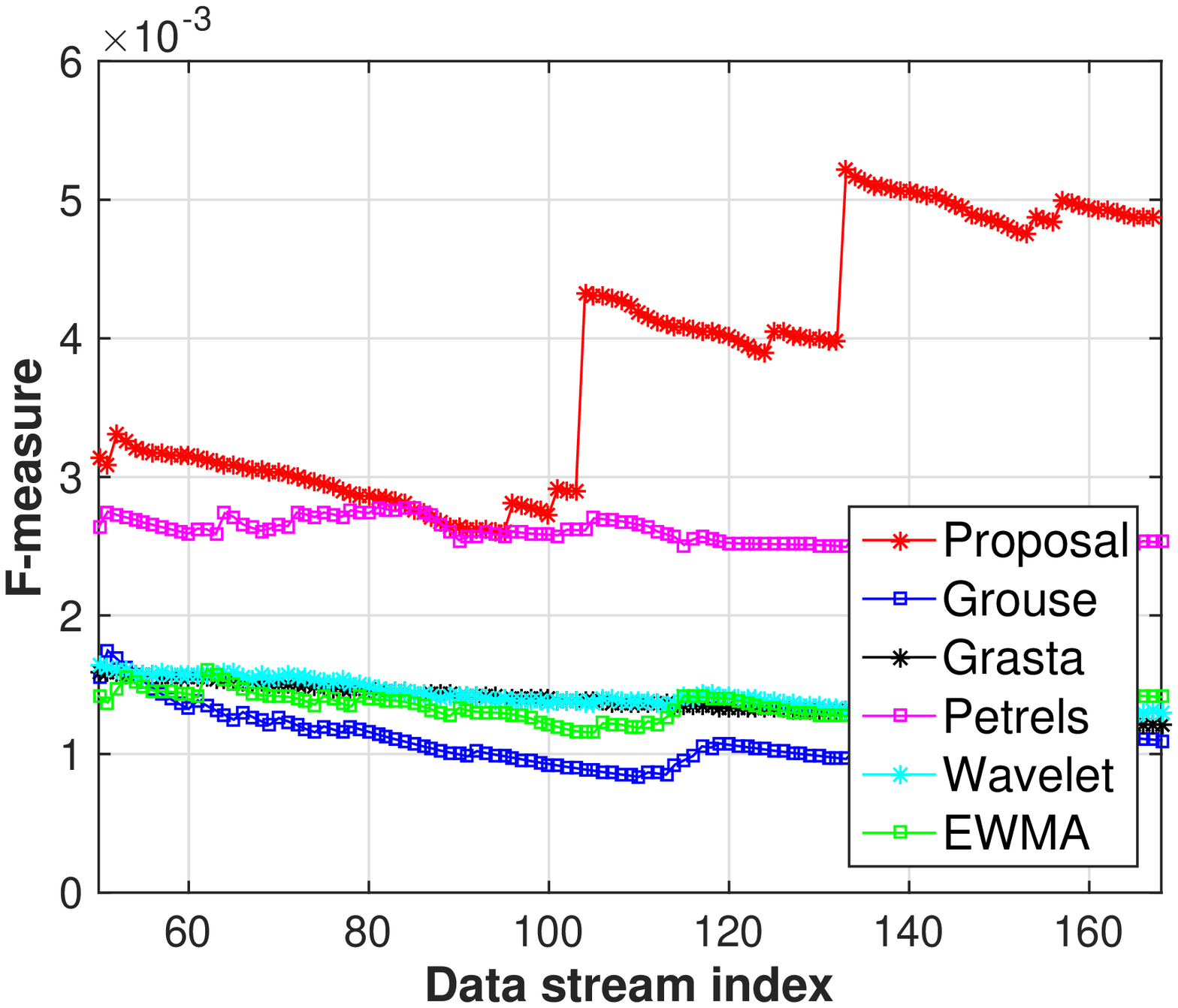}
        
         {(C) F-Measure}
        \end{center}
        
	\vspace*{-0.5cm} 
	\caption{Residual error and ROC (small-size network, $\rho=30$).}
    \label{fig:Synthetic_SmallSize_OR_30}
\end{figure}		
It should be firstly noted that the starting index of the proposed algorithm is delayed by $24$ data stream indices because it has the Hankel structure with the $W$-lagged data as seen in Fig.\ref{fig:Synthetic_SmallSize_OR_30}. Thus, while the first issue we observe is the convergence speed at the beginning of subspace algorithms, that of the proposed algorithm is the fastest as can be seen in Fig.\ref{fig:Synthetic_SmallSize_OR_30}(a). The next observation is that the proposed algorithm indicates the highest ROC values in the entire range as can be seen in Fig.\ref{fig:Synthetic_SmallSize_OR_30}(b). The F-measure value in Fig.\ref{fig:Synthetic_SmallSize_OR_30}(c) also yields the superior performance of the proposed algorithm. Furthermore, we evaluate the case with $\rho=50$, where the parameter configurations are the same as the case in $\rho=30$. The results in Fig.\ref{fig:Synthetic_SmallSize_OR_50} are very similar to the results with the observation ratio $\rho$ of $30$.
\begin{figure}[htbp]
        \begin{center}
        \includegraphics[width=0.5\hsize]{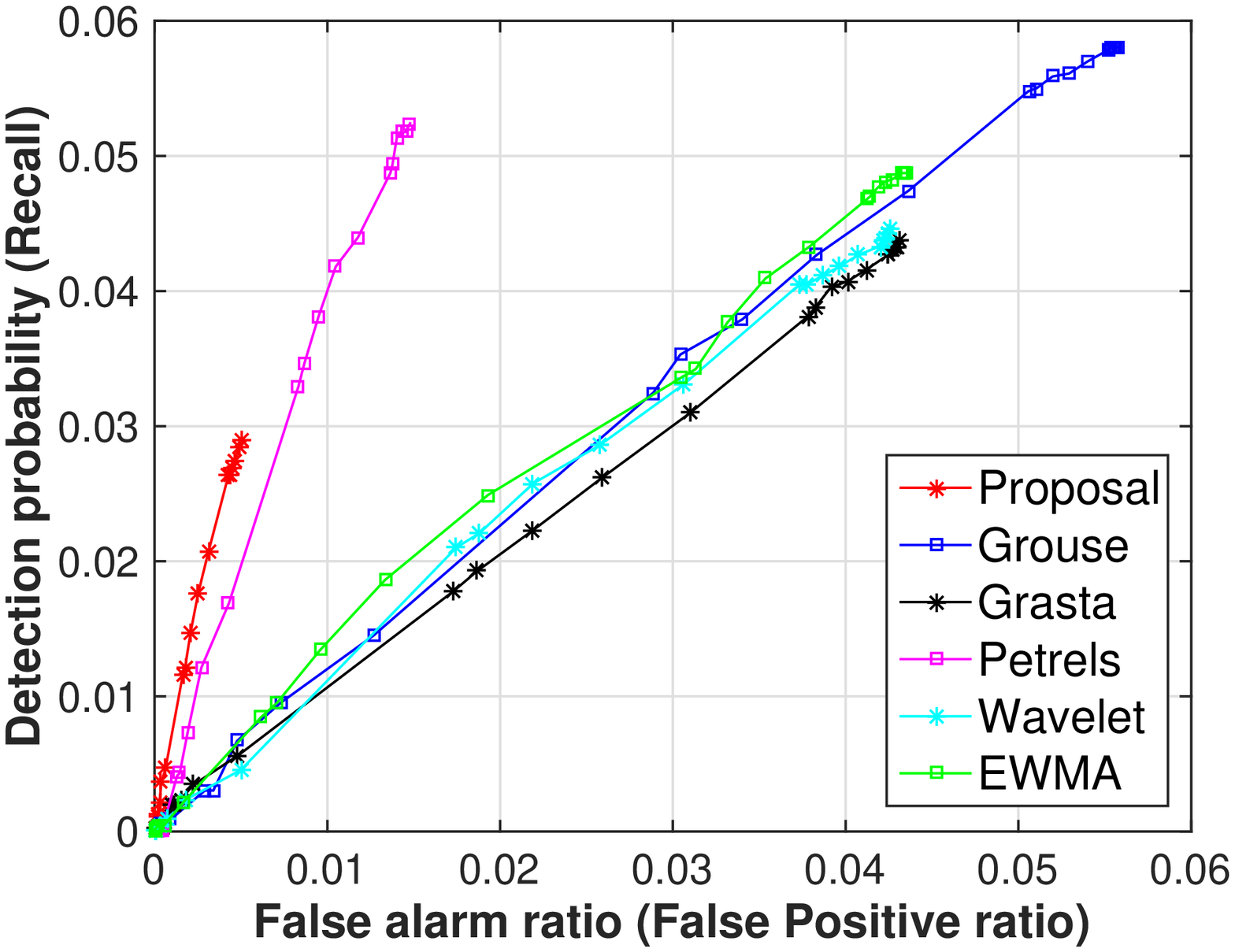}        
       
        \end{center}

    \caption{ROC (small-size network, $\rho=50$).}
    \label{fig:Synthetic_SmallSize_OR_50}
\end{figure}

\begin{figure}[htbp]
    \begin{tabular}{cc}
    \hspace*{-0.25cm}\begin{minipage}{0.5\hsize}
        \begin{center}
        \includegraphics[width=\hsize]{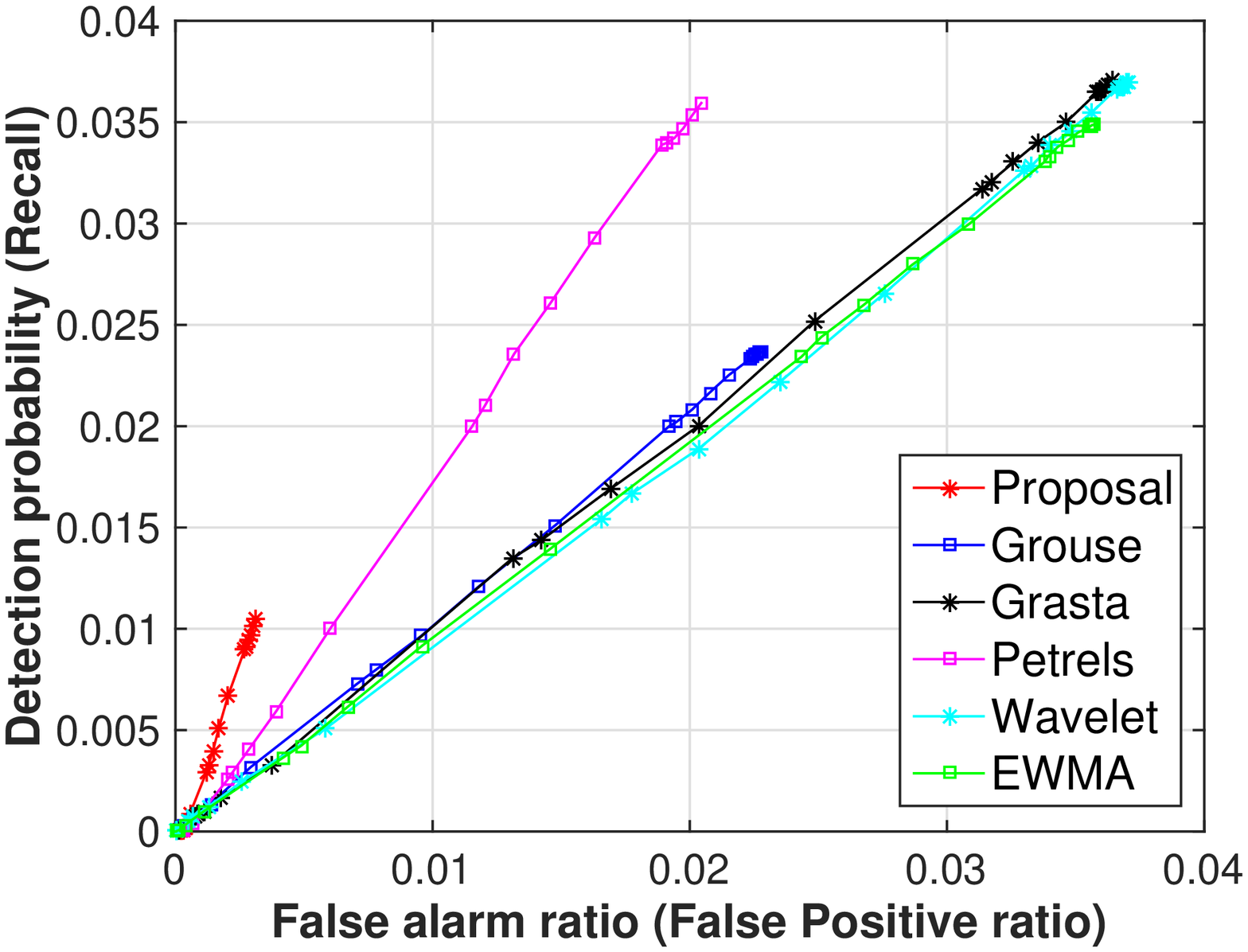}   
        
        {(a) $\rho=30$}
        \label{}
        \end{center}
    \end{minipage}
    \begin{minipage}{0.5\hsize}
        \begin{center}
        \includegraphics[width=\hsize]{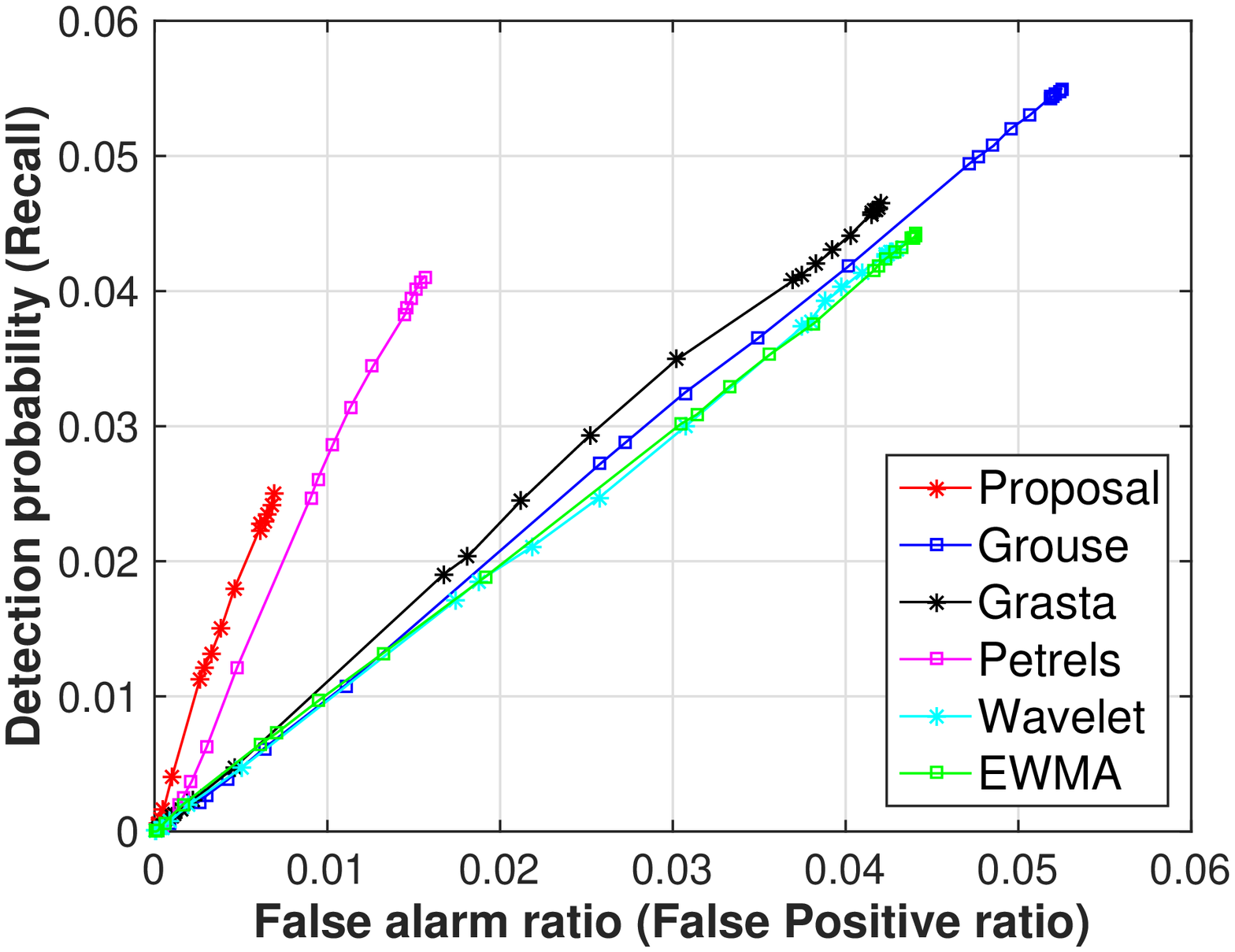}         
        
        {(b) $\rho=50$}
        \label{fig:fall}
        \end{center}
    \end{minipage}
    \end{tabular}
    \caption{ROC in biased flow structure (small-size network).}
    \label{fig:Synthetic_DifferentAnomalyType}
    \vspace*{0.5cm}

    \begin{tabular}{cc}
    \hspace*{-0.4cm}\begin{minipage}{0.5\hsize}
        \begin{center}
        \includegraphics[width=\hsize]{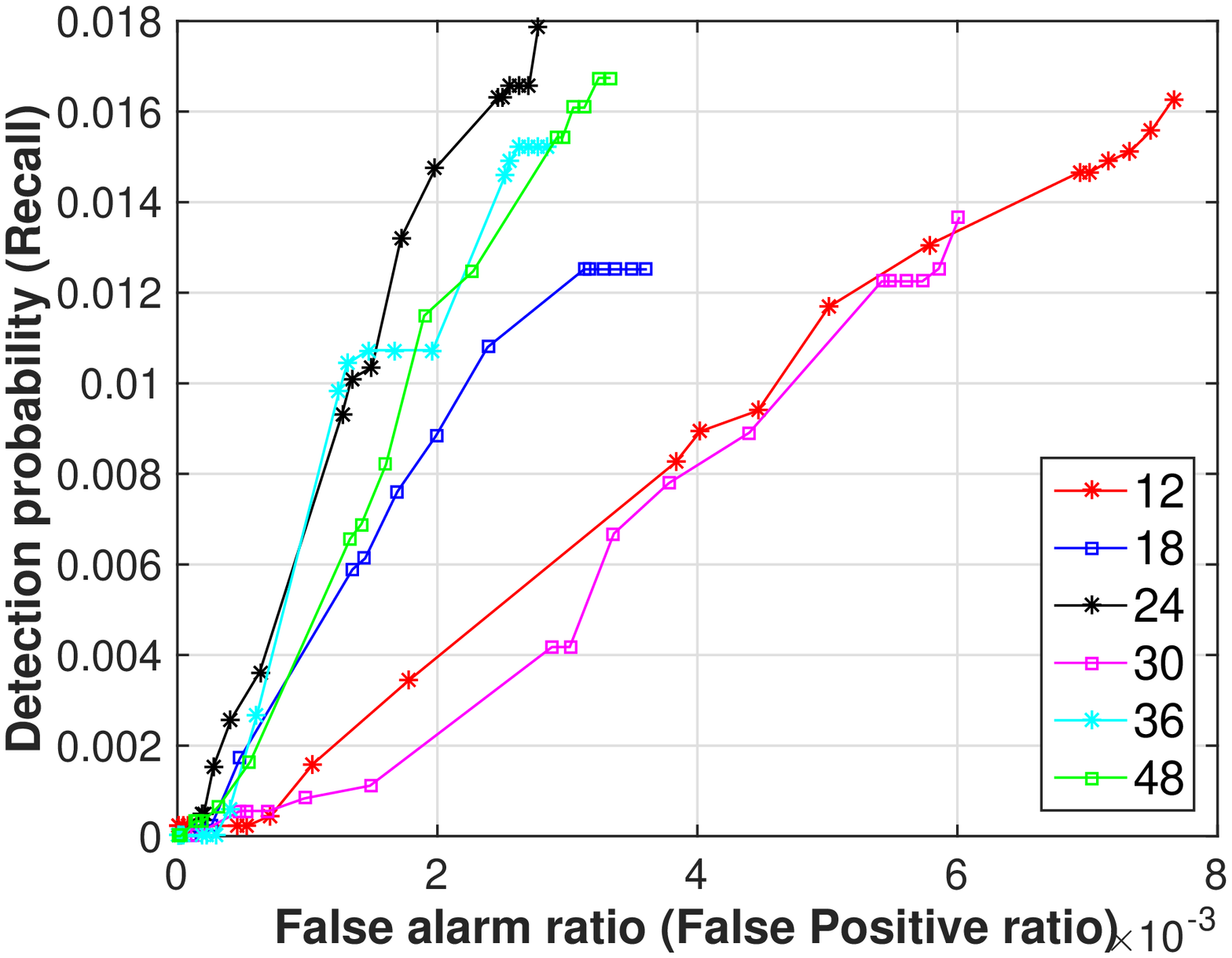} 
        
        {(a) $\rho=30$}
        \label{}
        \end{center}
    \end{minipage}
    \hspace*{0.1cm}\begin{minipage}{0.5\hsize}
        \begin{center}
        \includegraphics[width=\hsize]{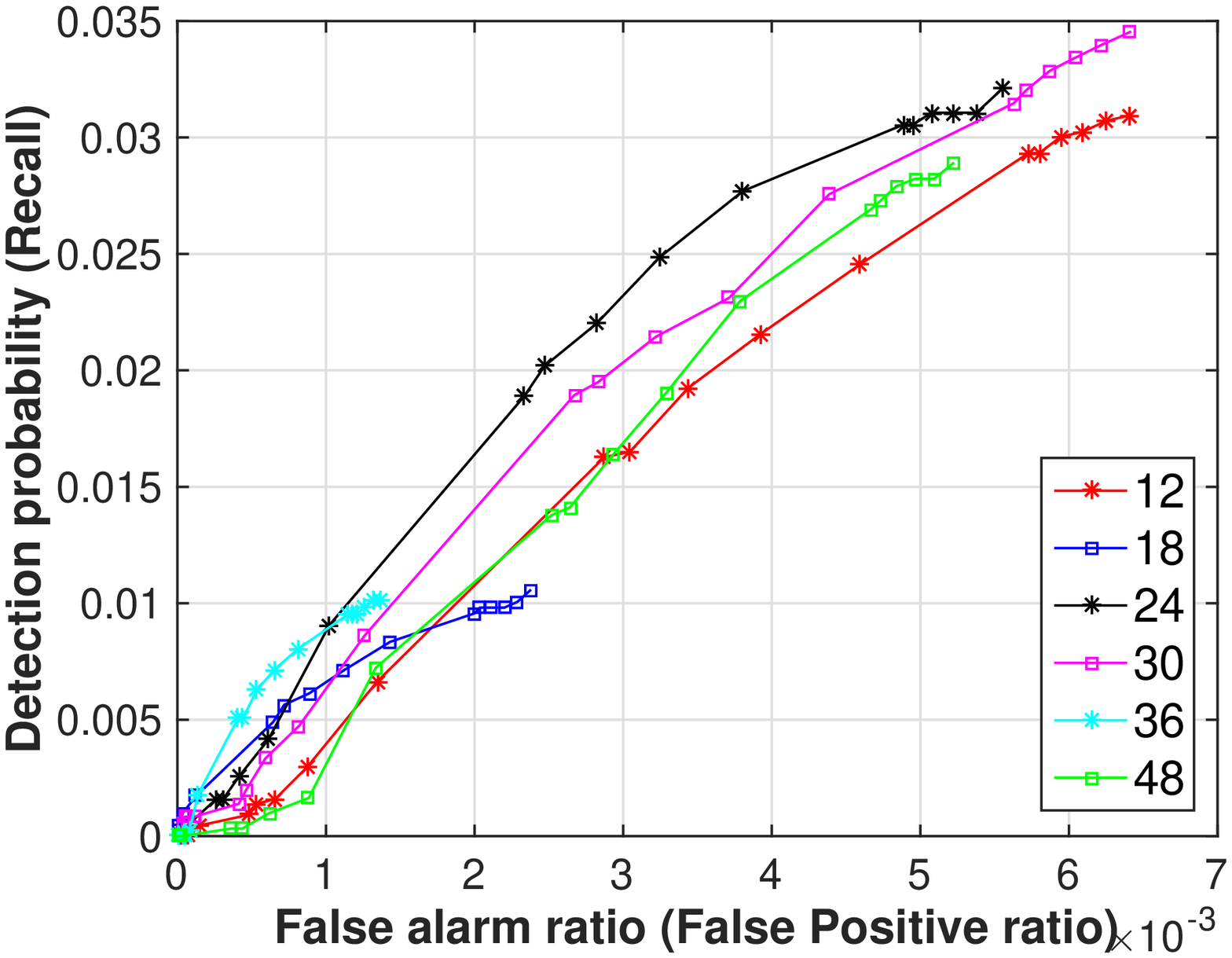}        
        
        {$\rho=50$}
        \label{fig:fall}
        \end{center}
    \end{minipage}
    \end{tabular}
    \caption{ROC (Different window lengths).}
    \label{fig:Synthetic_DifferentWindowLentgh}
\end{figure}

We also evaluate the behavior of the algorithms with varying types of the injected anomalies. This especially focuses on the difference between ``Mixture" case and ``Only N-1" case. The ``Mixture" case includes ``1-1", ``N-1", and ``All-ODs-one-link" types of abnormal flows in Table \ref{Tbl:SyntheticAnomalyInjectionCofig}. To the contrary, the ``Only N-1" case contains only ``N-1" types of abnormal flows. The other configurations are the same as the first evaluation experiments. The results when $\rho=30$ and $\rho=50$ are shown in Figs.\ref{fig:Synthetic_DifferentAnomalyType}(a) and (b), respectively. Although we cannot directly compare these with Fig.\ref{fig:Synthetic_SmallSize_OR_30}(b) and Fig.\ref{fig:Synthetic_SmallSize_OR_50} because the incomplete data position and the number and positions of anomaly flows are completely different, the results with the ``Only N-1" case of GROUSE, GRASTA, Wavelet and the EWMA are similar with the ``Mixture" case. Meanwhile, whereas PETRELS decreases largely,  the proposed algorithm remains the similar performance ($\rho=30$) or slightly decreases ($\rho=50$). This is due to the fact that the proposed algorithm can capture structure changes in particular. The abnormal flows in the ``Only N-1" case have a more biased structure than those in the ``Mixture" case, where the constituent flows are fairly distributed among nodes and links across the entire network.

Finally, the impact on the different window lengths $W=\{12,18,24,30,36,48\}$ for the Hankel structure are evaluated in Fig.\ref{fig:Synthetic_DifferentWindowLentgh}. The other configurations are the same as the first evaluation experiments. Fig.\ref{fig:Synthetic_DifferentWindowLentgh} reveals that $W=24$ yields the best performance in both $\rho=30$ and $\rho=50$ cases because the window length matches with the traffic periodicity.

\subsubsection{Experimental Results in Mid-size and Large-size Networks}

We further consider a mid-size network and a large-size network with the same number of flows $F=10^5$. For the mid-size network, $N_{node}$ is $1000$, $L$ is $11946$, and the ratio of injected anomalies is $1.59 \times 10^{-2}$\%. Meanwhile, the large-size network has $N_{node}$ is $3000$, $L$ is $17931$, and the ratio of injected anomalies is $1.93\times 10^{-2}$\%. Figs.\ref{fig:Synthetic_MiddleSize_OR_30} and  \ref{fig:Synthetic_LargeSize_OR_30} show the results for the mid-size and the large-size networks when $\rho$ is $30$, respectively. These figures yield the superior performance of the proposed algorithm against other algorithms.
		
\begin{figure}[htbp]
   \begin{tabular}{cc}
   \hspace*{-0.4cm}\begin{minipage}{0.5\hsize}
        \begin{center}
        \includegraphics[width=\hsize]{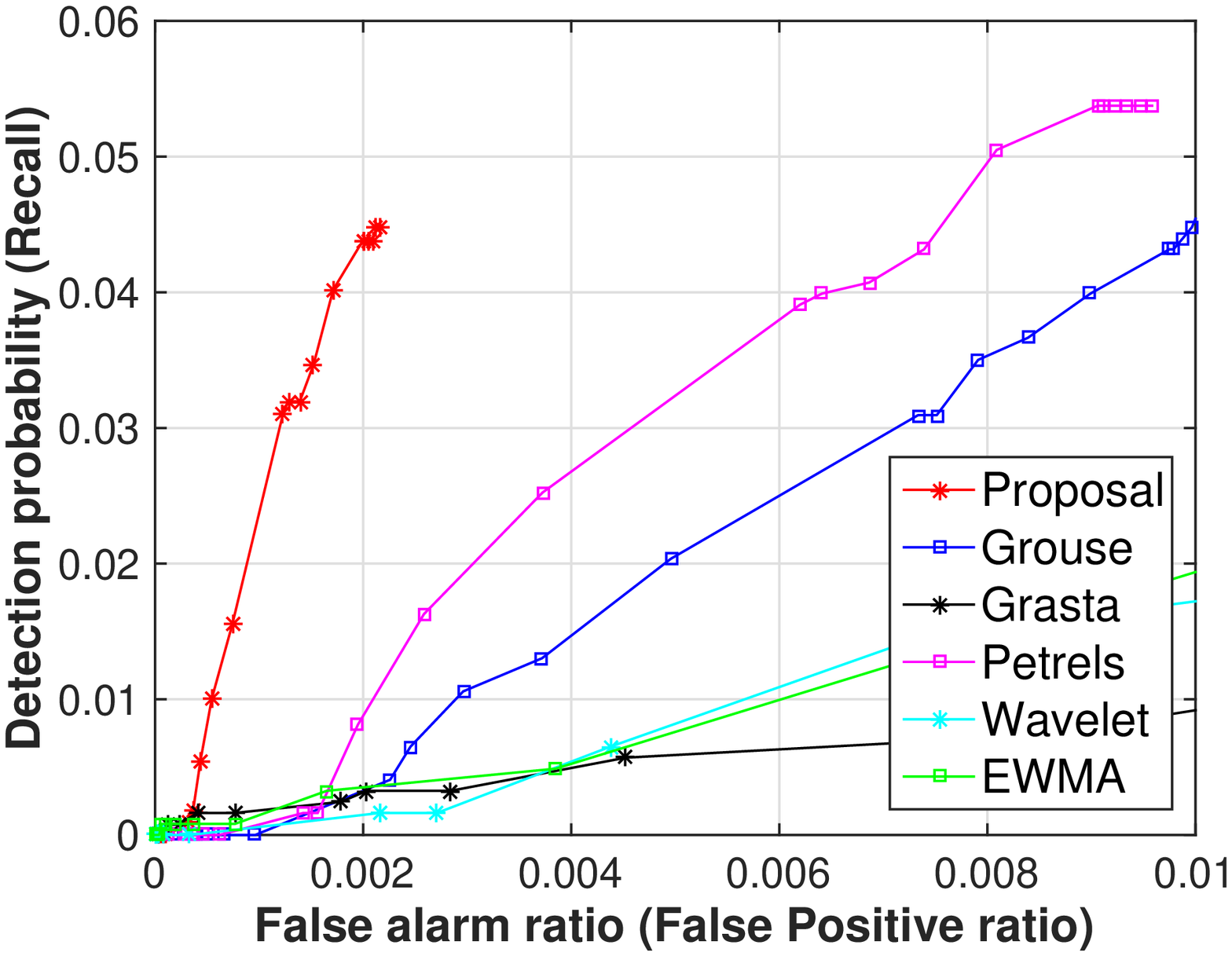}
        
        {(a) ROC}
        \label{}
        \end{center}
    \end{minipage}
    \hspace*{0.1cm}\begin{minipage}{0.5\hsize}
        \begin{center}
         \includegraphics[width=\hsize]{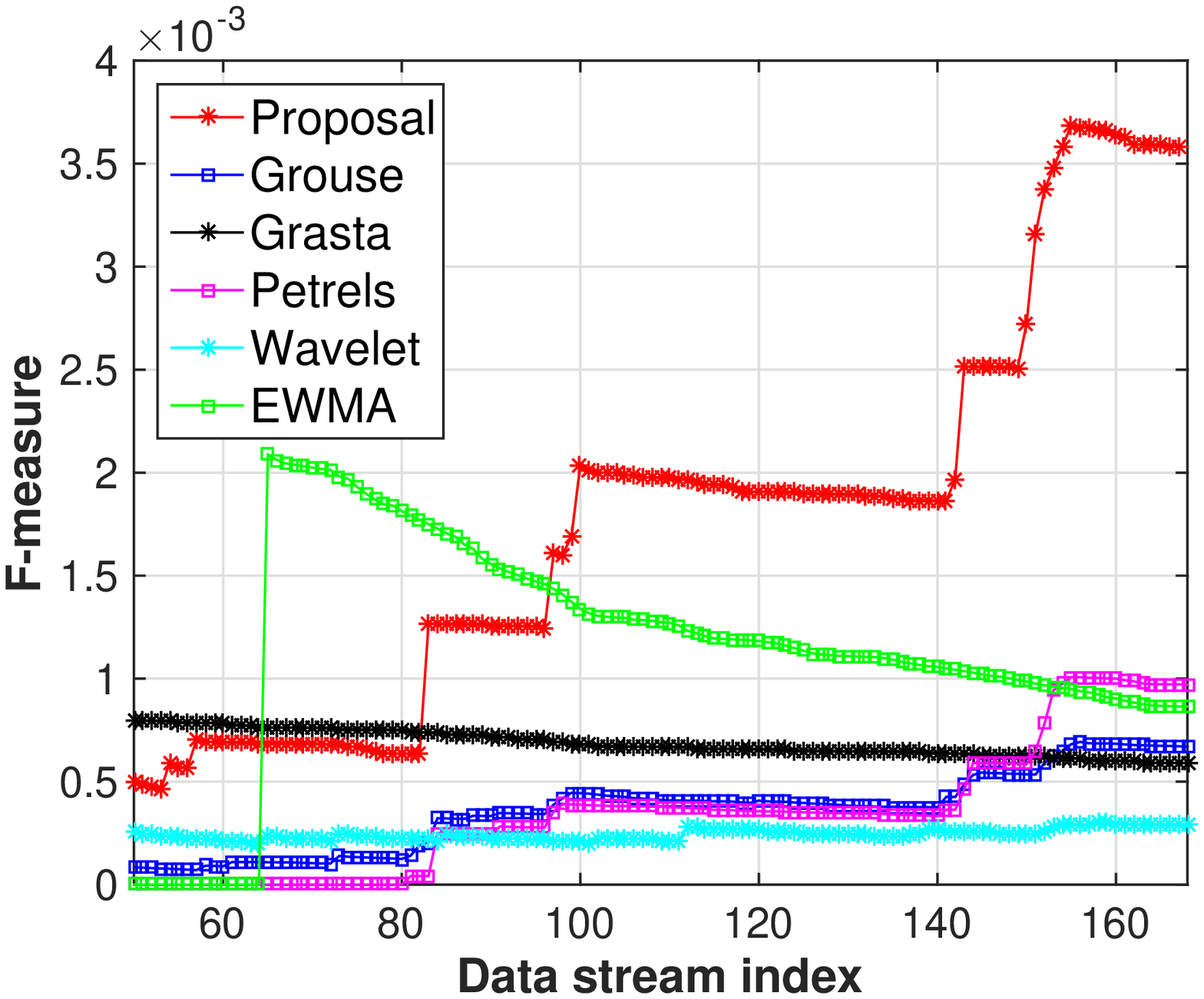}
        
        {(b) F-measure}
        \label{}
        \end{center}
    \end{minipage}
    \end{tabular}
    \caption{ROC and F-measure in mid-size network.}
    \label{fig:Synthetic_MiddleSize_OR_30}
    \vspace*{0.2cm}
 
   \begin{tabular}{cc}
   \hspace*{-0.4cm}\begin{minipage}{0.5\hsize}
        \begin{center}
        \includegraphics[width=\hsize]{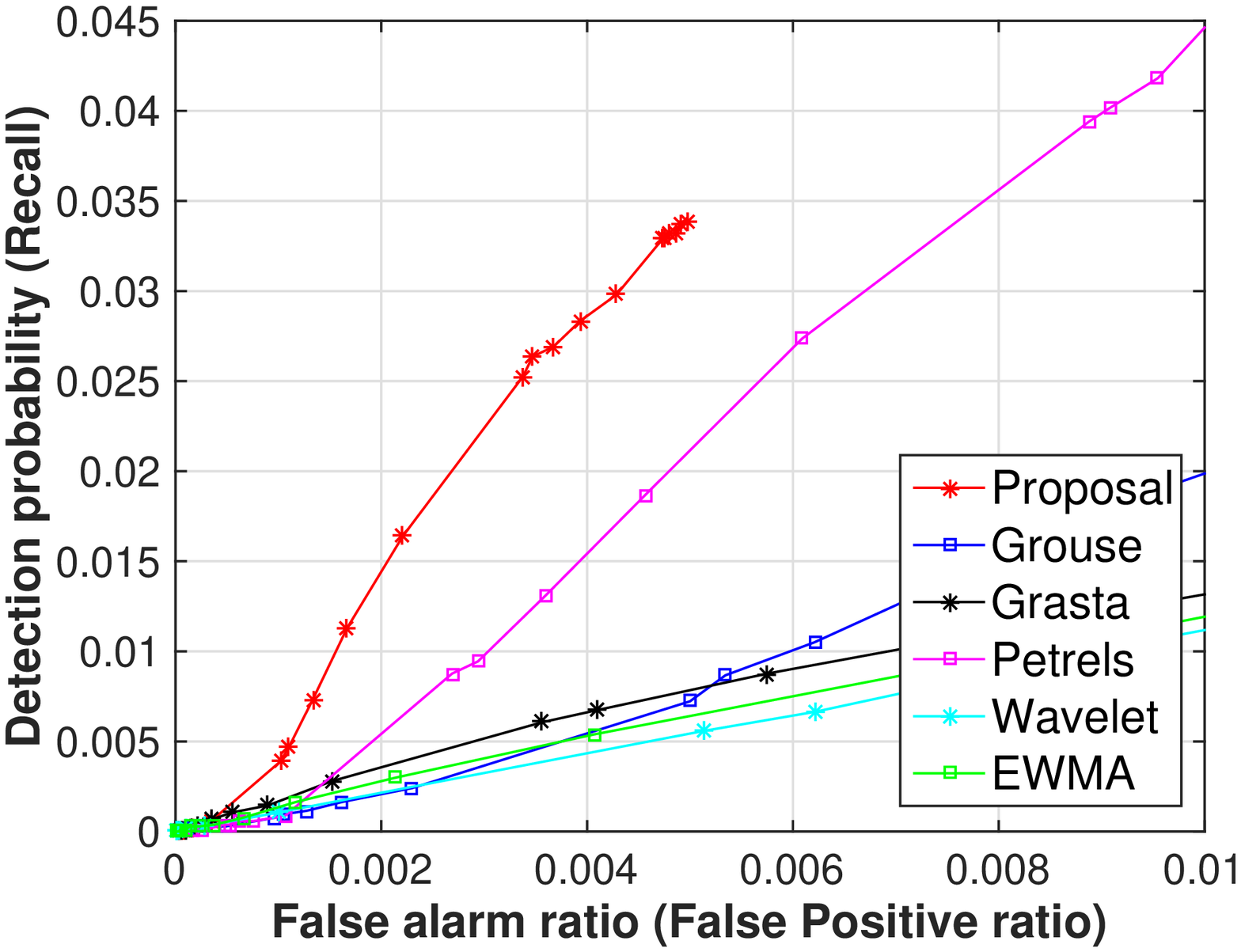}
        
        {(a) ROC}
        \label{}
        \end{center}
    \end{minipage}
    \hspace*{0.2cm}\begin{minipage}{0.5\hsize}
        \begin{center}
         \includegraphics[width=\hsize]{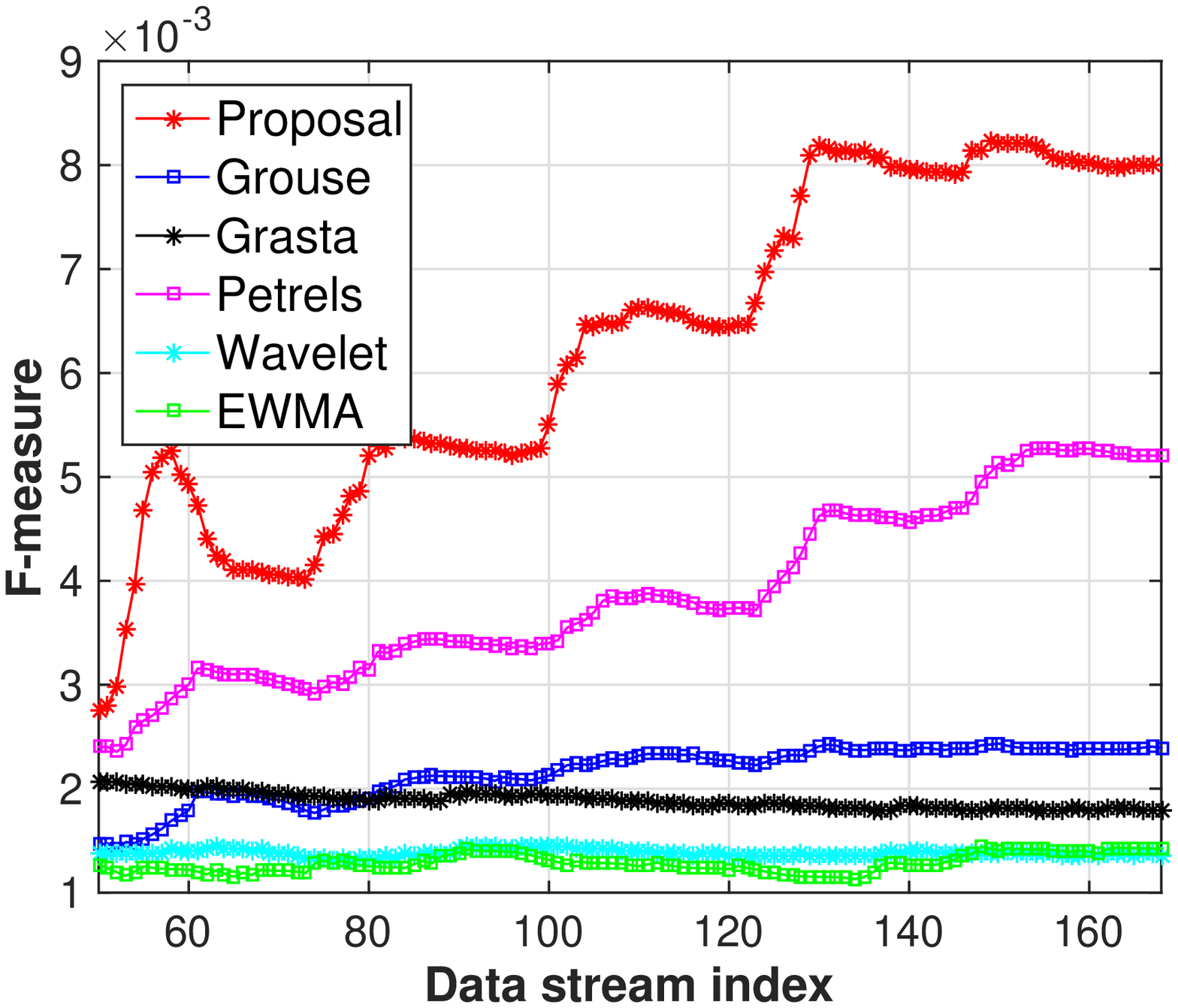}
        
        {(b) F-measure}
        \label{}
        \end{center}
    \end{minipage}
    \end{tabular}
    \caption{ROC and F-measure in large-size network.}
    \label{fig:Synthetic_LargeSize_OR_30}
\end{figure}

Finally, Table \ref{Tbl:ProcessingTime} shows the processing time for the mid-size network and the large-size network, respectively. The table shows not only the total processing time but also its breakdown, where ``A'' and ``B'' show the processing time for subspace tracking of $\{\vec{b}, \mat{A}, \mat{C}\}$, and the processing time for calculation of abnormal flows $\vec{v}$, respectively.
It should be noted that, whereas ``A'' depends on the implementation of each algorithm, ``B'' is calculated in the same implementation. The implemented code of our algorithm is not optimized. For the mid-size network, the proposed algorithm requires a longer tracking time (``A'') for multiple matrix inversions per iteration while a shorter calculation time (``B'') of sparse abnormal, \vec{v}, is needed compared to other algorithms. On the other hand, for the large-size network, the proposed algorithm is much faster. Since the subspace estimated by the proposed algorithm is much closer to the real subspace and is able to more efficiently capture the underlying time-series structure than others, the subsequent $\ell_1$ calculation of the proposed algorithm converges much faster than others. The results also reveals that the processing times almost match the computational complexity analysis in Section \ref{Sec:ComputationalComplexityAnalysis} with respect to the size of $L$ and $F$.

\begin{table}[htbp]
\begin{center}
\vspace*{0.3cm}
\caption{Processing time for mid-size and large-size network.}
\label{Tbl:ProcessingTime}
\begin{tabular}{l||r|r|r||r|r|r}
\hline
Network size & \multicolumn{3}{c||}{Middle (${\it N_{node}}=2000$)} & \multicolumn{3}{c}{Large  (${\it N_{node}}=3000$)}  \\
\hline
Algorithm& A \ \ & B \ \ & Total & A\ \  & B\ \ \ & Total  \\
\hline
\hline
Proposed &  393.1 & {\bf 1707.7} & {\bf 2100.8}  & 693.7 & {\bf 3585.9} & {\bf 4279.7} \\
\hline
GROUSE & 2.3 & 2339.5 & 2341.8 &   0.5 & 4688.5 & 4689.0  \\
\hline
GRASTA & 3.0 & 2347.3 & 2350.2 &  1.2 & 4676.8 & 4678.0    \\
\hline
PETRELS & 11.5 & 2341.4 & 2352.9 &  14.4 & 4674.9 & 4689.3\\
\hline
Wavelet &  33.8 & 2352.8 & 2386.6 &50.8 & 4667.6 & 4718.5   \\
\hline
EWMA &  {\bf 0.1} & 2336.6 & 2336.7 & {\bf 0.1} & 4626.9 & 4627.0\\
\hline
\end{tabular}
\end{center}
\end{table}

\section{Conclusion}

In this paper, we have addressed the challenge of detecting volume anomalies in large-scale data communication networks in an unsupervised way, where only the link traffic can be observed consisting of superimposed flows. For this purpose, the present paper assumes that anomalies in the flow can be identified by means of deviations of the measurements from a low rank structure. The network flow is modeled by means of a third order tensor with Hankel structure along one slice matrix to represent time-directional features as well as the spatial correlation of multiple links. By exploiting this traffic tensor with the Candecomp/PARAFC decomposition, a new online subspace tracking of the underlying low rank structure is proposed for normal flows based on the recursive least squares (RLS) method under partial observation. Meanwhile, abnormal flows are estimated as outlier sparse flows via sparsity maximization in the under-constrained linear-inverse problem.
An inherent shortcoming of our approach, which is shared by all unsupervised detection methods that are based on subspace tracking, is that anomalies which consist of change of the traffic at a very low frequency cannot be detected.
Numerical evaluations show that the proposed algorithm detects and identifies abnormal flows more accurately and with less computation time than the state-of-the-art online algorithms for a large-scale network. As future research directions, we plan to extend our studies to the cases where even the direct flow traffic data is only partially-observable. Additionally, the convergence analysis is a challenging task of the proposed online tensor optimization, and this remains an open problem.

\section*{Acknowledgement}
This work was initiated while H. Kasai was with the Department of Electrical and Computer Engineering (ECE) of Technical University of Munich, Germany. Part of this work is part of a project that has received funding from the European Research Council (ERC) under the European Unions Horizon 2020 research and innovation program (grant agreement No 647158 - FlexNets). (Corresponding author: Hiroyuki Kasai.)}

\section*{Appendix}

\appendix
\renewcommand\thefigure{A.\arabic{figure}}  
\setcounter{figure}{0} 

\renewcommand\thetable{A.\arabic{table}}  
\setcounter{table}{0} 

\renewcommand{\theequation}{A.\arabic{equation}}
\setcounter{equation}{0}

\section{Hankel matrix}
\label{Append_Sec:hankel}
Let $\vec{y} = \{y_{1}, \ldots ,y_{N}\}$ of length $N$ be a one-dimensional data. Given a window length $W$, with $1<W<N$, we construct the $W$-lagged vectors $\vec{h}_{k} = (y_{k}, \ldots , y_{k+W-1})^T \in \mathbb{R}^{W}$, $k = 1, \ldots, K$, where $K=N-W+1$, and compose these vectors into the Hankel matrix $\mat{H} = [\vec{h}_{1}: \cdots :\vec{h}_{K}] \in \mathbb{R}^{W \times K}$ as
\begin{eqnarray}  
   \mat{H} \ =\ [\vec{h}_1:\ldots:\vec{h}_K] \ =\ \left(
    \begin{array}{ccccc}
      y_{1} \!\!& y_{2} \!\!& y_{3} &\!\! \ldots \!\!& y_{K} \\
      y_{2} \!\!& y_{3} \!\!& y_{4} &\!\! \ldots \!\!& y_{K+1} \\
      \vdots & \vdots & \vdots & \!\!\ddots \!\!& \vdots \\
      y_{W} & y_{W+1} & y_{W+2} &\!\! \ldots \!\!& y_{N}
    \end{array}
  \right) . \nonumber
\end{eqnarray}   
This matrix is often called a {\it trajectory matrix}, which means that all the elements along the anti-diagonal are equal. 

\section{Candecomp/PARAFAC tensor decomposition}
\label{Append_Sec:CPDEC}

\begin{figure}[htbp]
	\begin{center}
	\includegraphics[width=14.5cm, bb=0 0 1220 387]{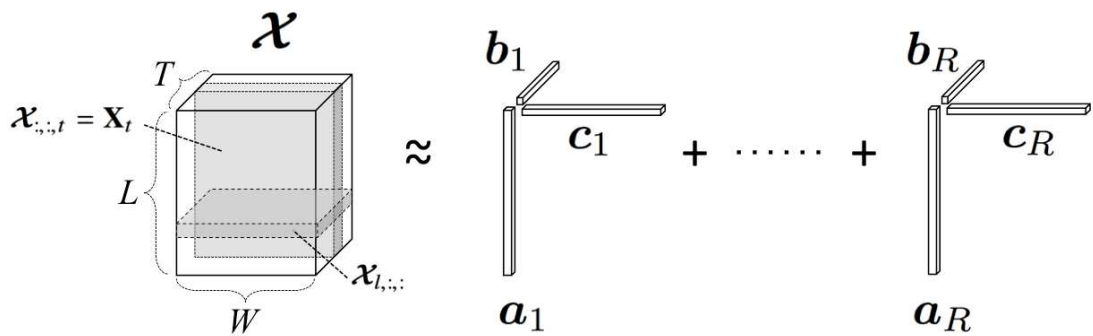}
	\caption{Candecomp/PARAFAC tensor decomposition.}
	\label{Append_Fig:BasicConcept}
	\end{center}
\end{figure}

The Candecomp/PARAFAC decomposition decomposes a tensor into a sum of component rank-one tensors \cite{Kolda_SIAMReview_2009}. Let $\mathbfcal{X}$ be a third-order tensor of size $L\times W \times T$, and assume its {\it rank} is $R$, we approximate $\mathbfcal{X}$ as $\mathbfcal{X}  \approx \sum_{r=1}^R \vec{a}_r \circ \vec{b}_r  \circ \vec{c}_r=\sum_{r=1}^R \vec{c}^t(r)\vec{a}_ r \vec{b}_r^T$, where $\vec{a}_r \in \mathbb{R}^L$, $\vec{b}_r \in \mathbb{R}^W$, and $\vec{c}_r \in \mathbb{R}^T$. The symbol ``$\circ$'' represents the vector outer product. Fig.\ref{Append_Fig:BasicConcept} illustrates rank-one tensor decomposition of the Candecomp/PARAFAC decomposition. The {\it factor matrices} refer to the combination of the vectors from the rank-one components, i.e., \mat{A} = $[\vec{a}_1: \vec{a}_2: \cdots : \vec{a}_R] \in \mathbb{R}^{L \times R}$ and likewise for $\mat{B} \in \mathbb{R}^{W \times R}$ and $\mat{C }\in \mathbb{R}^{T \times R}$. It should be emphasized that \mat{A}, \mat{B} and \mat{C} can be also represented by {\it row vectors}, i.e., {\it horizontal vectors}, namely, 
$\mat{A} = [(\vec{a}^1)^T: \cdots: (\vec{a}^L)^T]^T$, 
$\mat{B} = [(\vec{b}^1)^T: \cdots : (\vec{b}^T)^T]^T$, and
$\mat{C} = [(\vec{c}^1)^T : \cdots : (\vec{c}^W)^T]^T$, where $\{\vec{a}^l, \vec{b}^t,\vec{c}^w\} \in \mathbb{R}^{R}$. Thus, $\mat{X}_t\approx\mat{A}{\rm diag}(\vec{b}^t)\mat{C}^T$. 

\section{Derivation of {\bf A}[t] by Recursive Least Squares}
\label{Append_Sec:RLS}

This appendix describes the derivation of {\bf A}[t]. The derivative of (\ref{eq:problem_def_am}) with regard to $\vec{a}^l$ is calculated as 
\begin{eqnarray}
	\frac{\partial{F(\vec{a}^l)}}{\partial{(\vec{a}^l})} & = & 
	\sum_{\tau=1}^t 
	\biggl[-\sum_{w=1}^W  
	\lambda^{t-\tau} ({\bf \Omega}_{\tau})_{l,w}
	\left((\mat{Z}_\tau)_{l,w} 
	- (\vec{a}^l)^T \vec{\alpha}_w[\tau] \right)  \vec{\alpha}_{w} [\tau] \nonumber \\
	& &+\ \mu_h \sum_{w=1}^{W\!-\!1} \lambda^{t-\tau} ({\bf \Omega}_{\tau})_{l,w} 
	 (\vec{a}^l)^T \vec{\beta}_w[\tau] (\vec{\beta}_w[\tau])^T
	\biggr] 
	+ \mu_r[t]\vec{a}^l.\nonumber
\end{eqnarray}
Then, by setting this derivative equal to zero, we get $\vec{a}^l[t]$ as
\begin{equation}
\begin{split}
	&\left( \sum_{\tau=1}^t 
	\biggl[
	\sum_{w=1}^W  \lambda^{t-\tau} ({\bf \Omega}_{\tau})_{l,w} 
	\vec{\alpha}_w[\tau] (\vec{\alpha}_w[\tau])^T + \mu_h \sum_{w=1}^{W\!-\!1} \lambda^{t-\tau} ({\bf \Omega}_{\tau})_{l,w} 
	\vec{\beta}_w[\tau]   (\vec{\beta}_w[\tau])^T
	\biggr]+ \mu_r[t]\mat{I}_R \right)
	\vec{a}^l[t]\nonumber
	\\
	&\hspace*{2cm}= \sum_{\tau=1}^t 
	\sum_{w=1}^W  \lambda^{t-\tau}  ({\bf \Omega}_{\tau})_{l,w} (\mat{Z}_\tau)_{l,w} 
	 \vec{\alpha}_w[\tau] ,
\end{split}	 
\end{equation}
where $((\vec{a}^l)^T[t] \vec{\alpha}_w[\tau]) \vec{\alpha}_w[\tau]= ((\vec{\alpha}_w[\tau])^T \vec{a}^l[t]) \vec{\alpha}_w[\tau]
= \vec{\alpha}_w[\tau]((\vec{\alpha}_w[\tau])^T \vec{a}^l[t])$ is used. Finally, we get the following;
\begin{eqnarray}	 
	\mat{RA}_l[t] \vec{a}^l[t]  & = &  \vec{s}_l[t] \nonumber \\
	\vec{a}^l[t]  & =  & (\mat{RA}_l[t])^{\dagger} \vec{s}_l[t],\nonumber
\end{eqnarray}	
where  $\mat{RA}_l[t] \in \mathbb{R}^{R\times R}$ and $\vec{s}_l[t] \in \mathbb{R}^{R}$ are defined as 
\begin{eqnarray}	
 \mat{RA}_l[t] & :=&   \sum_{\tau=1}^t 
 	\biggl[
 	\sum_{w=1}^W  \lambda^{t-\tau} ({\bf \Omega}_{\tau})_{l,w} 
	\vec{\alpha}_w[\tau] (\vec{\alpha}_w[\tau])^T +
	\mu_h \sum_{w=1}^{W\!-\!1} \lambda^{t-\tau} ({\bf \Omega}_{\tau})_{l,w} 
	\vec{\beta}_w[\tau] (\vec{\beta}_w[\tau])^T 
	\biggr]+ \mu_r[t]\mat{I}_R, \nonumber \\
	\vec{s}_l[t] & := &  \sum_{\tau=1}^t 
	\sum_{w=1}^W  \lambda^{t-\tau}  ({\bf \Omega}_{\tau})_{l,w} (\mat{Z}_\tau)_{l,w} \vec{\alpha}_w[\tau].\nonumber
\end{eqnarray}	
Here, $\mat{RA}_l[t]$ is transformed as (\ref{appeq:Update_RA}).
\begin{figure*}[!t]
\begin{eqnarray}
	\label{appeq:Update_RA}
	\mat{RA}_l[t]  & = & 
	\sum_{\tau=1}^{t-1}
 	\left(
 	\sum_{w=1}^W  \lambda^{t-\tau} ({\bf \Omega}_{\tau})_{l,w} 
	\vec{\alpha}_w[\tau] (\vec{\alpha}_w[\tau])^T
	+
	\mu_h \sum_{w=1}^{W\!-\!1} \lambda^{t-\tau} ({\bf \Omega}_{\tau})_{l,w} 
	\vec{\beta}_w[\tau]  (\vec{\beta}_w[\tau])^T 
	\right)
	\nonumber \\
	&&+ \sum_{w=1}^W ({\bf \Omega}_t)_{l,w}  \vec{\alpha}_w[t]  (\vec{\alpha}_w[t])^T
	+ \mu_h  \sum_{w=1}^{W\!-\!1} ({\bf \Omega}_t)_{l,w}  \vec{\beta}_w[t]  (\vec{\beta}_w[t])^T 
	 + \mu_r[t]  \mat{I}_R
 	\nonumber \\
	&=& \lambda \biggl[
	\sum_{\tau=1}^{t-1}
 	\left(
 	\sum_{w=1}^W  \lambda^{t-1-\tau} ({\bf \Omega}_{\tau})_{l,w} 
	\vec{\alpha}_w[\tau] (\vec{\alpha}_w[\tau])^T
	+
	\mu_h \sum_{w=1}^{W\!-\!1} \lambda^{t-1-\tau} ({\bf \Omega}_{\tau})_{l,w} 
	\vec{\beta}_w[\tau] (\vec{\beta}_w[\tau])^T 
	\right) + \mu_r[\tau]  \mat{I}_R  \biggr]
	\nonumber \\
	&&+  \sum_{w=1}^W ({\bf \Omega}_t)_{l,w}  \vec{\alpha}_w[t]  (\vec{\alpha}_w[t])^T
	+ \mu_h  \sum_{w=1}^{W\!-\!1} ({\bf \Omega}_t)_{l,w}  \vec{\beta}_w[t]  (\vec{\beta}_w[t])^T 
 	+ \mu_r[t]  \mat{I}_R - \lambda \mu_r[t\!-\!1]  \mat{I}_R \nonumber \\
 	&=&  \lambda \mat{RA}_l[t\!-\!1]
	+  \sum_{w=1}^W ({\bf \Omega}_t)_{l,w}  \vec{\alpha}_w[t]  \vec{\alpha}_w[t]^T
	+ \mu_h  \sum_{w=1}^{W\!-\!1} ({\bf \Omega}_t)_{l,w} \vec{\beta}_w[t] \vec{\beta}_w[t]^T+ (\mu_r[t]  - \lambda \mu_r[t\!-\!1] ) \mat{I}_R.
\end{eqnarray}
\end{figure*}
Likewise, $\vec{s}_l[t]$ is also transformed as 
\begin{eqnarray}
	\vec{s}_l[t] 
	& = & 
	\sum_{w=1}^W \biggl[  
	\lambda^{t-1}  ({\bf \Omega}_1)_{l,w}  (\mat{Z}_{1})_{l,w} \vec{\alpha}_{w}[1]  + \cdots
	 +  \lambda^{1} ({\bf \Omega}_{t\!-\!1})_{l,w} (\mat{Z}_{t\!-\!1})_{l,w} \vec{\alpha}_{w}[t\!-\!1] + \lambda^{0} ({\bf \Omega}_t)_{l,w}  (\mat{Z}_t)_{l,w} \vec{\alpha}_w[t]  \biggr]\nonumber \\
	& = & \lambda \vec{s}_l[t\!-\!1] + \sum_{w=1}^W ({\bf \Omega}_t)_{l,w} (\mat{Z}_t)_{l,w} \vec{\alpha}_w[t]. \nonumber 
\end{eqnarray}
From $\mat{RA}_l[t] \vec{a}^l[t]  = \vec{s}_l[t] $, we modify this as (\ref{appeq:RA_a}). 
\begin{equation}
\begin{array}{lll}
	\label{appeq:RA_a}
	\mat{RA}_l[t] \vec{a}^l[t]  
	&=&   \lambda \mat{RA}_l[t\!-\!1]  \vec{a}^l[t\!-\!1] 
	 +\displaystyle{\sum_{w=1}^W ({\bf \Omega}_t)_{l,w} (\mat{Z}_t)_{l,w} \vec{\alpha}_w[t]} 
	\nonumber \\
	 &=&   \left(\mat{RA}_l[t] - \displaystyle{\sum_{w=1}^W ({\bf \Omega}_t)_{l,w}  \vec{\alpha}_w[t]  (\vec{\alpha}_w[t])^T} \right.  -\mu_h \displaystyle{\sum_{w=1}^{W\!-\!1}  ({\bf \Omega}_t )_{l,w+1}  \vec{\beta}_w[t]  (\vec{\beta}_w[t])^T} \\
	&&\Biggl.-  (\mu_r[t] - \lambda \mu_r[t\!-\!1]) \mat{I}_R   \Biggr) \vec{a}^l[t\!-\!1] + \!\!
	\displaystyle{\sum_{w=1}^W ({\bf \Omega}_t)_{l,w}  (\mat{Z}_t)_{l,w}   \vec{\alpha}_w[t] }
	\nonumber \\
	 &=&  \Biggl(\mat{RA}_l[t] 
	- \mu_h \displaystyle{\sum_{w=1}^{W-1}  ({\bf \Omega}_t)_{l,w+1}  \vec{\beta}_w[t]  (\vec{\beta}_w[t])^T} \Biggr. -  \Biggl. (\mu_r[t] - \lambda \mu_r[t-1]) \mat{I}_R \Biggr)
	 \vec{a}^l[t\!-\!1]  \nonumber \\
	&&+ \displaystyle{\sum_{w=1}^W ({\bf \Omega}_t)_{l,w} \left((\mat{Z}_t)_{l,w} - (\vec{\alpha}_w[t])^T  \vec{a}^l[t-1] \right) \vec{\alpha}_w[t]}.
\end{array}
\end{equation}

Finally, we obtain $\vec{a}^l[t]$ as (\ref{Eq:al_final}) by $\mat{Z}_\tau = \mat{Y}_\tau-{\mat{V}_{\scriptsize \mat{R}}}_\tau$.

\bibliographystyle{IEEEtran}
\bibliography{network_analysis,matrix_tensor_completion}

\end{document}